\documentclass[12pt]{article}
\pdfoutput=1
\usepackage{subfigure}
\usepackage{amssymb,amsmath}
\usepackage{graphicx}
\usepackage{color}
\usepackage[colorlinks=true
,urlcolor=blue
,citecolor=blue
,linkcolor=blue
,pagecolor=blue
,linktocpage=true
,pdfproducer=medialab
]{hyperref}
\usepackage[a4paper]{geometry}
\usepackage{placeins}
\usepackage[utf8]{inputenc}
\usepackage{cancel}
\usepackage{arydshln}
\makeatletter \renewcommand{\@dotsep}{10000} \makeatother

\def\alt{\lesssim}
\def\agt{\gtrsim}

\newcommand\prd[3]{{\it Phys.\ Rev.\ }{\bf D #1} (#2) #3}
\newcommand\prl[3]{{\it Phys.\ Rev.\ Lett.\ }{\bf #1} (#2) #3}

\newcommand\scalemath[2]{\scalebox{#1}{\mbox{\ensuremath{\displaystyle #2}}}}

\def\mgut{M_{{\rm GUT}}}

\usepackage{amsmath}

\newcommand{\bea}{\begin{eqnarray}}
\newcommand{\eea}{\end{eqnarray}}

\newcommand{\beq}{\begin{equation}}
\newcommand{\eeq}{\end{equation}}

 \makeatletter
\def\alt{\mathrel{\mathpalette\gl@align<}}
\def\agt{\mathrel{\mathpalette\gl@align>}}
\def\gl@align#1#2{\lower.6ex\vbox{\baselineskip\z@skip\lineskip\z@
\ialign{$\m@th#1\hfil##\hfil$\crcr#2\crcr\sim\crcr}}} \makeatother

\usepackage{longtable}

\begin{document}
%
\vspace*{1.0cm}

\begin{center}
{
\bf\Large
Sparticle Spectroscopy and Dark Matter in a $U(1)_{B-L}$ extension of MSSM
}
\\[5mm]
{\Large Waqas~Ahmed$^{\,a,}$} \footnote{E-mail: \texttt{waqasmit@nankai.edu.cn}},
{\Large Shabbar~Raza$^{\,b}$} \footnote{E-mail: \texttt{shabbar.raza@fuuast.edu.pk}},
{\Large Qaisar Shafi $^{\,c}$} \footnote{E-mail: \texttt{qshafi@udel.edu }}
{\Large Cem Salih Un$^{\,d,}$} \footnote{E-mail: \texttt{cemsalihun@uludag.edu.tr}}
{\Large Bin Zhu$^{\,e,f}$} \footnote{E-mail: \texttt {zhubin@mail.nankai.edu.cn}}
\vspace*{0.25cm}
\centerline{$^{a}$ \it
School of Physics, Nankai University, No.94 Weijin Road, Nankai District, Tianjin,China}
\centerline{$^{b}$ \it Department of Physics, Federal Urdu University of Arts, Science and Technology, Karachi 75300, Pakistan}
\centerline{$^{c}$ \it Bartol Research Institute, Department of Physics and Astronomy,}
\centerline{\it University of Delaware, Newark, DE 19716, USA}
\centerline{$^{d}$ \it Department of Physics, Bursa Uluda\~{g} University, 16059 Bursa, Turkey
}
\centerline{$^{e}$ \it Department of Physics, Yantai University, Yantai 264005, P. R. China}
\centerline{$^{f}$ \it Department of Physics, Chung-Ang University, Seoul 06974, Korea}

\vspace*{1.cm}

\vspace{-1.0cm} 
{\bf Abstract}
\end{center}
\baselineskip 18pt

We consider a class of SUSY models in which the MSSM gauge group is {{supplemented}} with a gauged $U(1)_{B-L}$ {{symmetry}} and a global $U(1)_{R}$ {{symmetry}}. {{This extension introduces }} only {{electrically}} neutral states, and {{the new}} SUSY partners effectively double the number of states in the neutralino sector {{that now includes a blino (from $B-L$) and singlino from a gauge singlet superfield}}. If the DM density is saturated by a LSP neutralino, the model yields quite a rich phenomenology depending on the DM {{composition}}. {{The LSP relic density constraint}} provides a lower bound on the stop and gluino masses of about {{3 TeV}} and 4 TeV respectively, which is testable in the near future collider experiments such as HL-LHC. {{The}} chargino mass {{lies between 0.24 TeV and about 2.0 TeV,  which can be tested based on the allowed decay channels}}. We also find $m_{\tilde{\tau}_{1}}\gtrsim 500$ GeV, and $m_{\tilde{e}},m_{\tilde{\mu}},m_{\tilde{\nu}^{S,P}} \gtrsim 1$ TeV. We identify chargino-neutralino coannihilation processes in the mass region $0.24 \,{\rm TeV} \lesssim m_{\tilde{\chi}_{1}^{0}}\approx m_{\tilde{\chi}_{1}^{\pm}}\lesssim 1.5$ TeV, and also coannihilation processes involving stau, selectron, smuon and sneutrinos {{for}} masses around 1 TeV. {{In addition, $A_{2}$ resonance solutions are found around 1 TeV, and  $H_{2}$ and $H_{3}$ resonance solutions are also shown around 0.5 TeV and 1 TeV . Some of the $A_{2}$ resonance solutions with $\tan\beta \gtrsim 20$ may be tested by the $A/H\rightarrow \tau^{+}\tau^{-}$ LHC searches.}}. While the relic density constraint excludes the bino-like DM, it is still possible to realize higgsino, singlino and blino-like DM {{for}} various mass scales. We show that all these solutions {{will}} be tested in future direct detection experiments such as LUX-Zeplin and Xenon-nT.

\thispagestyle{empty}

\newpage

\addtocounter{page}{-1}


\section{Introduction}

The Standard Model (SM) with massless neutrinos possesses an accidental global $U(1)_{B-L}$ symmetry \cite{tHooft:1976rip}. Turning this into a local $U(1)_{B-L}$ symmetry with no anomalies is readily achieved through the introduction of one right-handed neutrino per family, similar to the left-right symmetric models of Ref.~\cite{Pati:1974yy}. A supersymmetric extension of this scenario yields a Minimal Supersymmetric Standard Model [MSSM] supplemented by a $U(1)_{B-L}$ gauge symmetry. 
The spontaneous breaking of $U(1)_{B-L}$ with a Higgs field carrying two units of $B-L$ charge leaves unbroken a $Z_2$ symmetry which is precisely `matter' parity. This $Z_2$ symmetry coincides with the subgroup of $Z_4$, the center of $SO(10)$ (more precisely $Spin (10)$), and it is realized as an unbroken symmetry if the $SO(10)$ breaking is achieved with fields in the tensor representations \cite{Kibble:1982ae}. The presence of this unbroken $Z_2$ symmetry ensures that the lightest supersymmetric particle (LSP) is stable and if neutral, it is a viable dark matter candidate.
Motivated by the planned Run-3 at the LHC next year, we explore the low energy consequences of MSSM supplemented by $U(1)_{B-L}$ as well as a $U(1)$ R symmetry. A low scale breaking of $U(1)_{B-L}$ symmetry in SUSY models yields a rich phenomenology \cite{Abdallah:2018gjj}. There appears a $Z^{\prime}$ gauge boson together with its SUSY fermion partner ${\tilde B}^{\prime}$. The search for $Z^{\prime}$ such as this is a major focus of the recent and planned experiments \cite{Accomando:2013sfa,ATLAS:2017wce}. Our investigation is based on the model proposed in Ref.~\cite{Lazarides:2016ofd} where the $U(1)_{B-L}$ gauge symmetry is broken at tree level. A unique renormalizable superpotential is realized due to the presence of a global $U(1)_{R}$ symmetry. Among other things the model contains diphoton resonances as well as new dark matter candidates. We present our results for the collider.

The rest of the paper is organized as follows: In Section \ref{model} we present the salient features of the model including some of the main predictions. Section \ref{sec:scan} discusses scanning procedure and various constraints we have imposed. In Section \ref{spectrum} we will present the collider and dark matter results of the model, which include two new dark mark candidates not found in the MSSM. Our conclusions are summarized in Section \ref{conclusion}.

\section{The Model}\label{model}

The renormalizable superpotential of the MSSM with R-parity conservation possesses 
three global symmetries, namely baryon number $U(1)_B$, lepton number $U(1)_L$ and a R-symmetry $U(1)_R$, where, for simplicity, we ignore the tiny non-perturbative violation of $B$ and $L$ by the $SU(2)_L$ instantons. We supplement the gauge symmetry of MSSM with local $U(1)_{B-L}$ symmetry whose spontaneous breaking is achieved as described in \cite{Khalil:2007dr}. The Higgs mechanism includes three new super fields $S$, $\Phi$ and $\bar{\Phi}$ which are singlets under the MSSM gauge group. The full renormalizable and gauge invariant superpotential can be written as

\bea
W &=&y_u \hat{H_{u}}\hat{q}\hat{u}^c + y_d \hat{H_{d}} \hat{q}\hat{d}^c + y_{\nu}\hat{H_{u}}\hat{l}\hat{\nu}^c +y_{e}\hat{H_{d}}\hat{l}\hat{e}^c
\nonumber
\\
& & +\kappa S (\Phi\bar{\Phi}-M^2) +\lambda_{\mu} SH_{u}H_{d}+ 
\lambda_{\nu^c}\bar\Phi \nu^c\nu^c 
\label{W}
\eea
where $y_u$, $y_d$, $y_{\nu}$, $y_e$ are the Yukawa couplings
and the family indices  are generally suppressed for simplicity. Also, $q$, $u^c$, $d^c$, $l$, $\nu^c$, $e^c$ are the usual quark and lepton 
superfields of MSSM including the right handed neutrinos $\nu^c$, and 
$H_{u}$, $H_{d}$ are the standard electroweak Higgs superfields. The $\Phi$ and $\bar{\Phi}$ fields carry non-zero $B-L$ charges, and their VEVs spontaneously break the $U(1)_{B-L}$ symmetry. {{ Furthermore, $\Phi$ is responsible for generating a Majorana mass term for the right-handed neutrinos. Such terms preserve the $B-L$ symmetry only if $\Phi(\bar \Phi)$ has $Q_{B-L}=2(-2)$. Since $\Phi$ and $\bar \Phi$ have twice the lepton number, they are called as bileptinos.}} The superfield $S$ only carries a $U(1)_{R}$ charge, and develops a non-zero VEV after soft SUSY breaking proportional to the gravitino mass ($m_{3/2}$). This can provide a resolution to the MSSM $\mu-$problem \cite{Dvali:1994wj}, since the $\mu-$term is effectively obtained as $\mu\equiv \lambda_{\mu}\langle v_{S} \rangle$. Table \ref{tab:fields} summarizes the quantum numbers of the various superfields. In addition, $S$, $\Phi$ and $\bar{\Phi}$ lead also to significant implications for inflationary phenomenology \cite{Rehman:2018nsn}.


\begin{table}
\centering 
\begin{tabular}{|c|c|c|c|c|c|c|} 
\hline \hline 
Matter Superfields & Spin 0 & Spin \(\frac{1}{2}\) & Generations & \(G_{SM} \otimes\, U(1)_{B-L}\) & R \\

\hline 
\(\hat{q}\) & \(\tilde{q}\) & \(q\) & 3 & \((\frac{1}{6},{\bf 2},{\bf 3},\frac{1}{6}) \) & 1\\ 
\(\hat{l}\) & \(\tilde{l}\) & \(l\) & 3 & \((-\frac{1}{2},{\bf 2},{\bf 1},-\frac{1}{2}) \)& 1\\ 
\(\hat{d}\) & \(\tilde{d}_R^{0,*}\) & \(d_R^*\) & 3 & \((\frac{1}{3},{\bf 1},{\bf \overline{3}},-\frac{1}{6}) \)& 1\\ 
\(\hat{u}\) & \(\tilde{u}_R^{0,*}\) & \(u_R^*\) & 3 & \((-\frac{2}{3},{\bf 1},{\bf \overline{3}},-\frac{1}{6}) \)& 1\\ 
\(\hat{e}\) & \(\tilde{e}_R^*\) & \(e_R^*\) & 3 & \((1,{\bf 1},{\bf 1},\frac{1}{2}) \)& 1\\ 
\(\hat{\nu}\) & \(\tilde{\nu}_R^*\) & \(\nu_R^*\) & 3 & \((0,{\bf 1},{\bf 1},\frac{1}{2}) \)& 1 \\ 
\hline
Higgs Superfields & & & & &\\
\hline
\(\hat{H}_d\) & \(H_d\) & \(\tilde{H}_d\) & 1 & \((-\frac{1}{2},{\bf 2},{\bf 1},0) \)& 0\\ 
\(\hat{H}_u\) & \(H_u\) & \(\tilde{H}_u\) & 1 & \((\frac{1}{2},{\bf 2},{\bf 1},0) \)& 0\\ 
\(\hat{\bar \Phi}\) & \(\bar \phi\) & \(\tilde{\bar \phi}\) & 1 & \((0,{\bf 1},{\bf 1},-1) \) & 0\\ 
\(\hat{\Phi}\) & \(\phi\) & \(\tilde{\phi}\) & 1 & \((0,{\bf 1},{\bf 1},1) \) & 0\\ 
\(\hat{S}\) & \(\tilde{S}\) & \(S\) & 1 & \((0,{\bf 1},{\bf 1},0) \)& 2\\ 
\hline \hline
\end{tabular}
\caption{Chiral superfields and their charges under the local guage symmetry $G_{SM}\times U(1)_{B-L}$ and $U(1)$ $R$-symmetry.}
\label{tab:fields} 
\end{table} 

The relevant soft supersymmetry breaking (SSB) terms are given as follows:

\begin{align} 
- L_{SB} =\,& L_{MSSM} +\tilde{S} T_{S} + H_d^0 H_u^0 \tilde{S} T_{\lambda_{\mu}} +H_d^- H_u^+ \tilde{S} T_{\lambda_{\mu}} -{{\phi}} {{\bar \phi}} \tilde{S} T_{\kappa}  +{\bar {\phi}} \tilde{\nu}^*_{R,{i}} \tilde{\nu}^*_{R,{k}} T_{\phi,{i k}} \nonumber \\
&-H_u^+ \tilde{\nu}^*_{R,{i}} \tilde{e}_{L,{j}} T_{\nu,{i j}}+H_u^0 \tilde{\nu}^*_{R,{i}} \tilde{\nu}_{L,{j}} T_{\nu,{i j}}+m_{\phi}^2 |\phi|^2 +m_{\bar{\phi}}^2 |\bar{\phi}|^2 +m_S^2 |\tilde{S}|^2 \nonumber \\
 & +\tilde{\nu}^*_{L,{i}} m_{l,{i j}}^{2} \tilde{\nu}_{L,{j}}+\tilde{\nu}^*_{R,{i}} m_{\nu,{i j}}^{2} \tilde{\nu}_{R,{j}}+\lambda_{\tilde{B}} {\tilde{B}{}'} {M}_{B B'}  + \frac{1}{2}{\tilde{B}{}'}^{2} {M}_{BL} \delta_{i j} + \mbox{h.c.}
\end{align} 
where $T_{S}=L_{0}\kappa M^{2}$, $T_{\lambda_{\mu}}=\lambda_{\mu} A_{\lambda_{\mu}}$, $T_{\kappa}=\kappa A_{\kappa}$, $T_{\lambda_{\nu^c}}=\lambda_{\nu^c}A_{0}$, $T_{\nu}=Y_{\nu}A_{0}$, and $i,j$ are generation indices.

As stated earlier, even though the chargino sector remains intact, {{the neutralino sector is enriched, since the fermionic components of $S$ (singlino), $\Phi$ and $\bar{\Phi}$ (bileptinos) can mix with the MSSM neutralinos after the symmetry breaking}}. In addition, the superpartner of $Z'$ is also allowed to mix with the other neutralinos. Thus the neutralino mass matrix becomes an $8\times 8$ matrix and it can be constructed in the basis $(\tilde{B},\tilde{W}^{0},\tilde{H}_{d}^{0},\tilde{H}_{u}^{0},\tilde{B}^{\prime},\tilde{\Phi}.\tilde{\bar{\Phi}},\tilde{S})$ as 

\begin{equation} 
m_{\tilde{\chi}^0} = \scalemath{0.9}{
\left(
\begin{array}{cccccccc}
M_1 &0 &-\frac{1}{2} g_1 v_d  &\frac{1}{2} g_1 v_u  &{M}_{B B'} &- g_{B Y} v_{\bar \phi}  &g_{B Y} v_{{\phi}}  &0\\ 
0 &M_2 &\frac{1}{2} g_2 v_d  &-\frac{1}{2} g_2 v_u  &0 &0 &0 &0\\ 
-\frac{1}{2} g_1 v_d  &\frac{1}{2} g_2 v_d  &0 &- \frac{1}{\sqrt{2}} \lambda v_s  &-\frac{1}{2} g_{Y B} v_d  &0 &0 &- \frac{1}{\sqrt{2}} \lambda v_u \\ 
\frac{1}{2} g_1 v_u  &-\frac{1}{2} g_2 v_u  &- \frac{1}{\sqrt{2}} \lambda v_s  &0 &\frac{1}{2} g_{Y B} v_u  &0 &0 &- \frac{1}{\sqrt{2}} \lambda v_d \\ 
{M}_{B B'} &0 &-\frac{1}{2} g_{Y B} v_d  &\frac{1}{2} g_{Y B} v_u  &{M}_{BL} &- g_{B} v_{\bar \phi}  &g_{B} v_{{\phi}}  &0\\ 
- g_{B Y} v_{\bar \phi}  &0 &0 &0 &- g_{B} v_{\bar \phi}  &0 &m_{\tilde{\phi}\tilde{\bar \phi}} &m_{S\tilde{\bar \phi}}\\ 
g_{B Y} v_{{\phi}}  &0 &0 &0 &g_{B} v_{{\phi}}  &m_{\tilde{\bar \phi}\tilde{\phi}} &0 &m_{S\tilde{\phi}}\\ 
0 &0 &- \frac{1}{\sqrt{2}} \lambda v_u  &- \frac{1}{\sqrt{2}} \lambda v_d  &0 &m_{\tilde{\bar \phi}S} & m_{\tilde{\phi}S} & 0
\end{array} 
\right)}
 \end{equation} 
where
\begin{equation}
m_{\tilde{\bar \phi}\tilde{\phi}} = - \frac{1}{\sqrt{2}} \lambda_2 v_s,\hspace{0.3cm} 
m_{\tilde{\bar \phi}S} = - \frac{1}{\sqrt{2}} \lambda_2 v_{{\phi}},\hspace{0.3cm}
m_{\tilde{\phi}S} = - \frac{1}{\sqrt{2}} \lambda_2 v_{\bar \phi}
\label{neutralinomass}
\end{equation}

After diagonalizing the mass matrix given in Eq.(\ref{neutralinomass}) as $m_{\tilde{\chi}^{0}}^{{\rm diag}}=N^{*}m_{\tilde{\chi}^{0}}N^{\dagger}$, the neutralino mass eigenstates can be obtained as

\begin{equation}
m_{\tilde{\chi}_{i}^{0}}=N_{ij}\lambda_{j},\hspace{0.3cm} i,j=1,\ldots,8,
\end{equation},
with
\begin{align} 
{\tilde{B}} = \sum_{j}N^*_{j 1}\lambda^0_{{j}}\,, \hspace{1cm} 
\tilde{W}^0 = \sum_{j}N^*_{j 2}\lambda^0_{{j}}\,, \hspace{1cm} 
\tilde{H}_d^0 = \sum_{j}N^*_{j 3}\lambda^0_{{j}}\\ 
\tilde{H}_u^0 = \sum_{j}N^*_{j 4}\lambda^0_{{j}}\,, \hspace{1cm} 
{\tilde{B}{}'} = \sum_{j}N^*_{j 5}\lambda^0_{{j}}\,, \hspace{1cm} 
\tilde{\phi} = \sum_{j}N^*_{j 6}\lambda^0_{{j}}\\ 
\tilde{\bar \phi} = \sum_{j}N^*_{j 7}\lambda^0_{{j}}\,, \hspace{1cm} 
S = \sum_{j}N^*_{j 8}\lambda^0_{{j}}
\end{align} 
where $\tilde B$, $\tilde W^{0}$, $\tilde{H}_d^0$, $\tilde{H}_u^0$, ${\tilde{B}{}'}$, $\tilde{\phi}$,
$\tilde{\bar {\phi}}$ and $S$ stand for bino, wino, higgsinos, blino, bileptinos and 
singlino receptively. The LSP neutralino can be written as follows:
\begin{equation}
\chi_{1}^{0}=N_{11} \tilde B +N_{12} \tilde W^{0} +N_{13} \tilde{H}_d^0 + N_{14} \tilde{H}_u^0 +N_{15}{\tilde{B}{}'} +N_{16} \tilde{\phi}+N_{17}\tilde{\bar \phi}+N_{18} S.
\end{equation}
{{In addition to bino-dominated LSP, the variety of neutralinos can also result in higgsino-like, singlino-like and blino-like LSPs, despite the universal gaugino mass term at the GUT scale.}}

\section{Phenomenological constraints and scanning procedure}\label{sec:scan}

We have employed SPheno 3.3.3 package \cite{Porod:2003um} generated with SARAH 4.9.0 \cite{Staub:2008uz}. In this package, the weak scale values of the gauge and Yukawa couplings are evolved to the unification scale $M_{{\rm GUT}}$ via the renormalization group equations (RGEs). $M_{{\rm GUT}}$ is determined by the requirement of the gauge coupling unification, described as $g_{3}\approx g_{2}=g_{1}=g_{B-L}$, where $g_{3}$, $g_{2}$ and $g_{1}$ are the MSSM gauge couplings for $SU(3)_{C}$, $SU(2)_{L}$ and $U(1)_{Y}$ respectively, while $g_{B-L}$ corresponds to the gauge coupling for $U(1)_{B-L}$. Concerning the contributions from the threshold corrections to the gauge couplings at $\mgut$ arising from some unknown breaking mechanisms of the GUT gauge group, $g_{3}$ receives the largest contributions \cite{Hisano:1992jj}, and it is allowed to deviate from the unification point up to about $3\%$. If a solution does not satisfy this condition within this allowance, SPheno does not generate an output for such solutions by default. Hence, the existence of an output file guarantees that the solutions {are} compatible with the unification condition, and $g_{3}$ deviates no more than $3\%$. With the boundary conditions given at $M_{{\rm GUT}}$, all of the SSB parameters along with the gauge and Yukawa couplings are evolved back to the weak scale.

We have performed random scans over the following parameter space:

\begin{minipage}{0.5\textwidth}
\begin{equation*}
\begin{array}{l}
0\leq m_{0} \leq 5 ~\rm{TeV}   ~,\\
0\leq  M_{1/2}  \leq 5 ~\rm{TeV}  ~, \\
4\leq M_{Z^{\prime}}  \leq 5 ~\rm{TeV}  ~,\\
0.3 \leq  v_{S}  \leq 20 ~\rm{TeV}  ~, \\
2\leq  \tan\beta  \leq 60   ~, \\
1\leq  \tan\beta^{\prime}  \leq 2~,
\end{array}
\end{equation*} 
\end{minipage}%
\begin{minipage}{0.5\textwidth}
\vspace{0.6cm}\begin{equation}
\begin{array}{l}
-3\leq  A_{0}/m_{0}  \leq 3   ~, \\
-3\leq  A_{\lambda_{\mu}}/m_{0}  \leq 3   ~,\\
-3\leq  A_{\kappa}/m_{0}  \leq 3   ~,\\
-3\leq  L_{0}/m_{0}  \leq 3   ~, \\
-2\leq  \lambda_{\mu}  \leq 2 ~, \\
-2\leq  \kappa  \leq 2   ~, \\
1\leq  L_{3}  \leq 10^{5} ~\rm{TeV}   ~,
\end{array}
\end{equation}
\end{minipage}
where $L_{3} = \kappa M^{2}$. 

We take the Yukawa matrix,
$\lambda_{\nu^{c}}$, between the right-handed neutrinos and the MSSM singlet Higgs field $\Phi$ as a diagonal matrix, with the entries given as $\lambda_{\nu^{c},11}=\lambda_{\nu^{c},22}=\lambda_{\nu^{c},33}\simeq 0.4$. Note that $\lambda_{\nu^{c}}$ is determined at the low scale, and values larger than about $0.4$ can yield  Landau pole below the GUT scale \cite{OLeary:2011vlq}. Also for simplicity, we set $Y_{\nu}=0$. Finally, we set the top quark mass to its central value ($m_{t} = 173.3$ GeV)
\cite{Group:2009ad}. Note that the sparticle spectrum is not {very}
sensitive to one or two sigma variation in the top quark mass
\cite{Gogoladze:2011db}, but it can shift the Higgs boson mass by
$1-2$ GeV  \cite{Gogoladze:2011aa}.

One of the important theoretical constraints arises from {radiative electroweak symmetry breaking} (REWSB) \cite{Ibanez:1982fr} by requiring $m_{H_{u}}\neq m_{H_{d}}$ at the low scale. Since we set $m_{H_{u}}=m_{H_{d}}$ at the GUT scale, the required separation between the masses of the MSSM Higgs fields can be generated through the RGEs with arbitrary large Yukawa coupling for top quark. In our case, radiative symmetry breaking is also employed for the $U(1)_{B-L}$ symmetry, and it constrains the Yukawa couplings in the $B-L$ sector.   Another important constraint comes from the relic abundance of the stable charged particles \cite{Nakamura:2010zzi},
which excludes regions where the charged SUSY particles such as stau and stop become the LSP. In our scans, we allow only solutions for which one of the neutralinos is the LSP and REWSB condition is satisfied. 
In scanning the parameter space, we use our interface, which employs
Metropolis-Hasting algorithm described in \cite{Belanger:2009ti}. Furthermore, we apply the following {constraints successively}.

\textbf{\underline{LEP constraints:}}
We impose the bounds that the LEP2 experiments set on charged sparticle masses ($\gtrsim 100$ GeV) \cite{Patrignani:2016xqp}.

\textbf{\underline{Higgs Boson mass:}}
The experimental combination for the Higgs mass reported by 
the ATLAS and CMS Collaborations is \cite{Khachatryan:2016vau}
\begin{align}\label{eqn:mh}
m_{h} = 125.09 \pm 0.21(\rm stat.) \pm 0.11(\rm syst.)~GeV .
\end{align}  
Due to an estimated 2 GeV theoretical uncertainty in the calculation of $m_h$ in the MSSM -- see {\it e.g.}~\cite{Allanach:2004rh} --
we apply the constraint from the Higgs boson mass to our results as follows
\begin{align}\label{eqn:higgsMassLHC}
123~ {\rm GeV} \leq m_h \leq 127~ {\rm GeV}. 
\end{align}

\textbf{\underline{Rare B-meson decays:}} Since the SM predictions are in a good agreement with the experimental results for the rare decays of $B-$meson such as the $B_{s}\rightarrow \mu^{+}\mu^{-}$, $B_{s}\rightarrow X_{s}\gamma$, where $X_{s}$ is an appropriate state including a strange quark, the results of our analyses are required to be consistent with the measurements for such processes. Thus we employ the following constraints from B-physics \cite{CMS:2014xfa,Amhis:2014hma}:

\begin{align}\label{eqn:Bphysics}
1.6\times 10^{-9} \leq ~ {\rm BR}(B_s \rightarrow \mu^+ \mu^-) ~
  \leq 4.2 \times10^{-9} ,\\ 
2.99 \times 10^{-4} \leq  ~ {\rm BR}(b \rightarrow s \gamma) ~
  \leq 3.87 \times 10^{-4}, \\
0.70\times 10^{-4} \leq ~ {\rm BR}(B_u\rightarrow\tau \nu_{\tau})~
        \leq 1.5 \times 10^{-4}.
\end{align}

\textbf{\underline{Current LHC searches:}}
Based on \cite{Aaboud:2017vwy,Vami:2019slp,Sirunyan:2017kqq}, we consider the following constraints on gluino and first/second generation squark masses
\begin{align}
(a) \quad \quad m_{\widetilde g} \gtrsim ~ 2.2 ~ {\rm TeV},\quad \quad m_{\widetilde q} \gtrsim ~ 2 ~ {\rm TeV},
\label{lhc-a}
\end{align}

Searches for two and three leptons plus missing energy \cite{Sirunyan:2017lae,Aaboud:2018jiw} set bounds on the electro-weak production of charged-neutral higgsinos decaying to $WZ$ and the LSP, which can be approximately translated into the following condition \cite{Ahmed:2017jxl}
\begin{align}
(b) \quad \quad {\rm if} ~~ m_{\widetilde \chi^{0}_{1}} < ~ 100 ~ {\rm GeV} ~\Longrightarrow ~ \tilde{\mu} > 350 ~ {\rm GeV}.
\label{lhc-b}
\end{align}

Finally the bound on $M_{Z^{\prime}}$ is being constantly updated by comprehensive analyses. The severest bound on $M_{Z'}$ comes from negative results from the LEP data as $M_{Z'}/g_{B-L} \geq 6$ TeV \cite{Cacciapaglia:2006pk}. Even though considering its decay modes can lower the mass bound on $Z^{\prime}$ \cite{Accomando:2013sfa}, setting $M_{Z^{\prime}} \geq 4$ TeV \cite{ATLAS:2017wce} guarantees avoiding possible exclusions on our solutions due to the light $Z^{\prime}$ mass. Thus we restrict its range in our scan as

\begin{equation}
(c) \quad \quad 4 \leq M_{Z^{\prime}} \leq 5~{\rm TeV}
\end{equation}

The upper bound on $M_{Z^{\prime}}$ in our scan is only for keeping its mass range in the testable era of the current and near future experiments.

\textbf{\underline{DM searches and relic density:}} For the discussion on the phenomenology of neutralino DM in our scenario, we impose the following constraint for the LSP relic density, based on {the current measurements of the Planck satellite} \cite{Akrami:2018vks}:
\begin{align}\label{eq:omega}
0.114 \leq \Omega_{\rm CDM}h^2 (\rm Planck2018) \leq 0.126   \; (5\sigma).
\end{align}

We use the current LUX and XENON1T spin-independent (SI) DM cross section with bounds as a constraint \cite{Akerib:2016vxi,Aprile:2017iyp}. All points lying above these upper bounds have been excluded from the plots. We also show the projection of future limits of already published of XENON1T with 2 $t\cdot y$ exposure and XENONnT with 20 $t\cdot y$ exposure \cite{Aprile:2015uzo}. Bounds for  spin-dependent cross section of DM set by
the current LUX and future LUX-ZEPLIN experiments \cite{Akerib:2017kat,Akerib:2016lao} are also shown.

\section{Sparticle Spectroscopy and Dark Matter}{\label{spectrum}}

In this section we are focusing on the sparticle spectrum consistent with mass bounds and the constraints discussed above.

\begin{figure}[htp!]
\centering
\subfiguretopcaptrue

\subfigure{
\includegraphics[totalheight=5.5cm,width=7.cm]{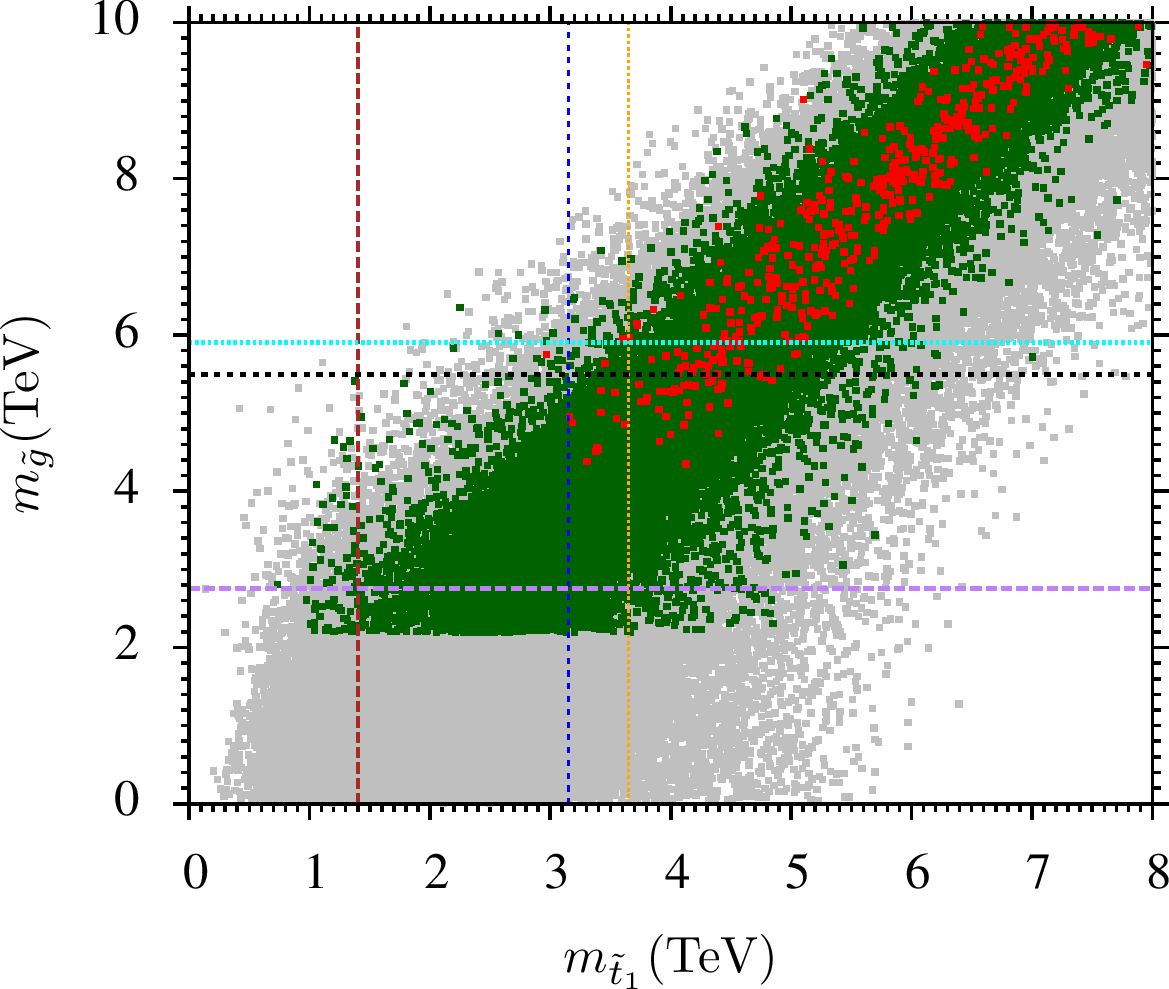}
}
\subfigure{
\includegraphics[totalheight=5.5cm,width=7.cm]{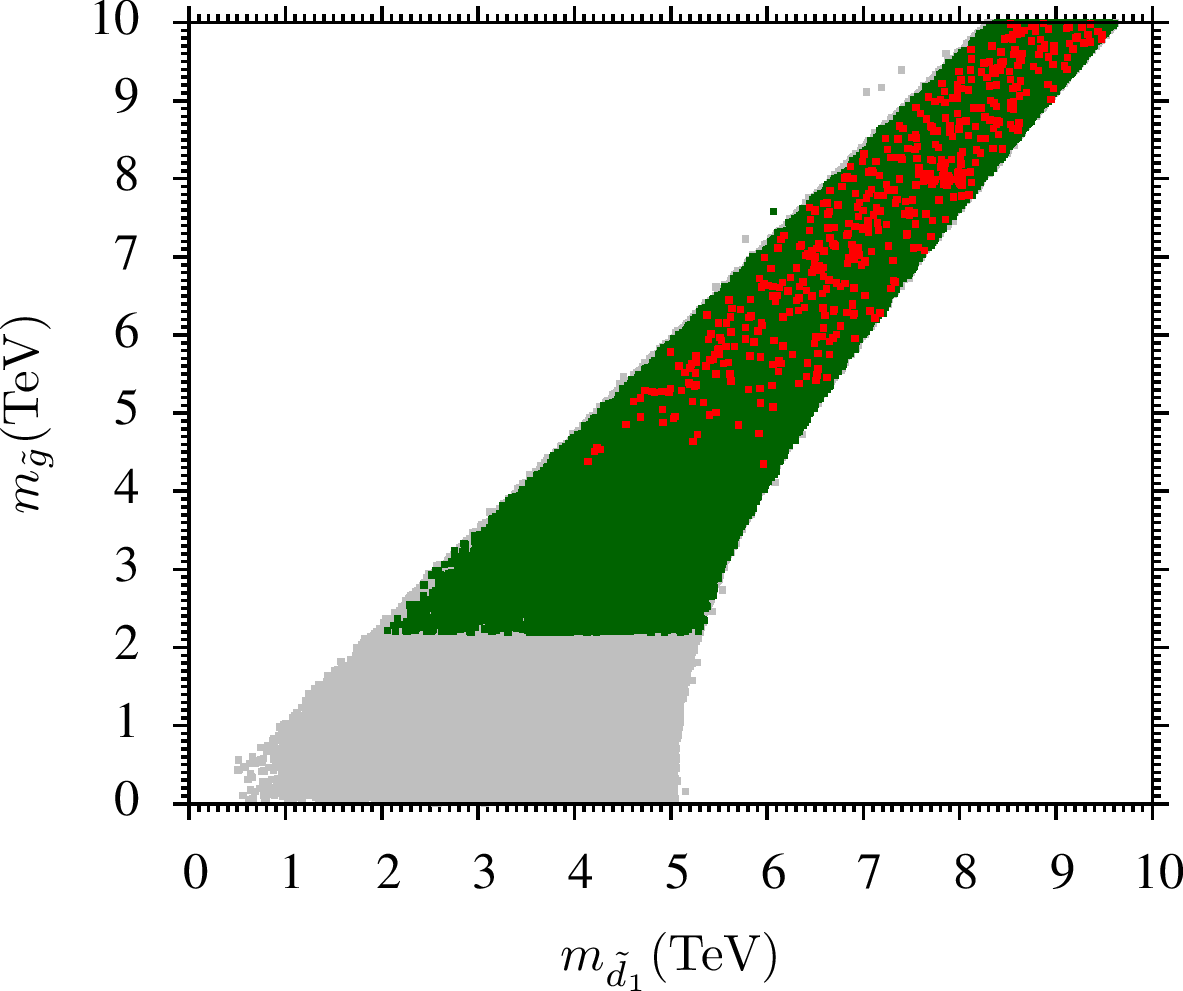}
}\\
\subfigure{
\includegraphics[totalheight=5.5cm,width=7.cm]{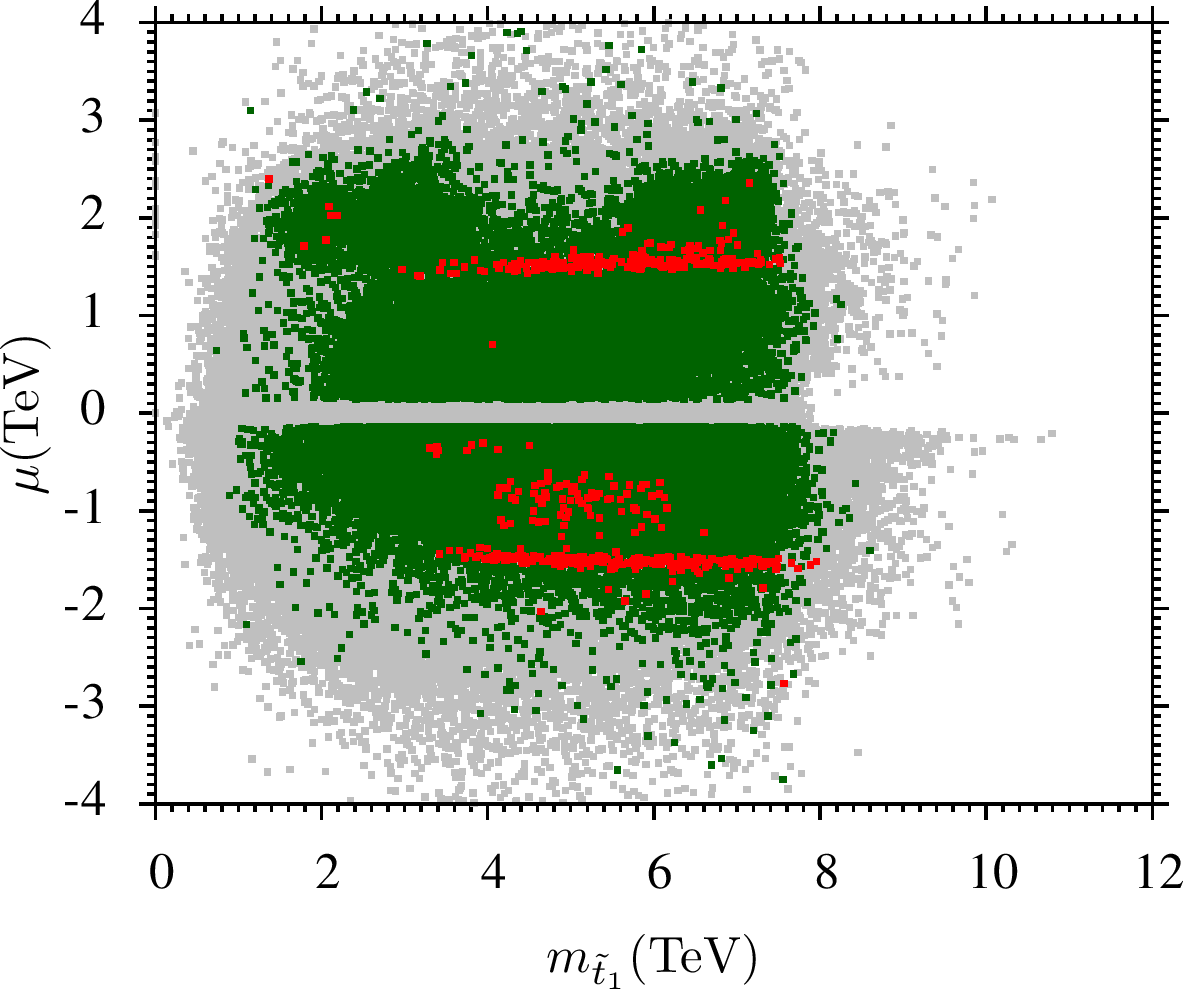}
}
\caption{Grey points satisfy the REWSB and the LSP neutralino conditions. Green points satisfy the mass bounds and B-physics constraints. Red points form a subset of green points and satisfy Planck2018 bounds on the relic abundance of the LSP neutralino within $5\sigma$ uncertainty. They are consistent with the current dark matter direct detection bounds presented by LUX. See text for the description of dashed horizontal and vertical lines in left panel.}

\label{gluino-stop}
\end{figure}

In Figure~\ref{gluino-stop}, we show plots in the $m_{\tilde t_{1}}-m_{\tilde g}$ and $m_{\tilde t_{1}}-\mu$ planes. Grey points satisfy the REWSB and the LSP neutralino conditions. Green points satisfy the mass bounds and B-physics constraints. Red points form a subset of green points and satisfy the Planck2018 bounds on the relic abundance of the LSP neutralino within $5\sigma$ uncertainty. They are also consistent with the current dark matter direct detection bounds presented by LUX. We see in our present scans that, the gluino mass can be between 2.2 to 10 TeV, while $m_{\tilde t_{1}}$ can be  from 1 to 8 TeV (green points). If we impose the DM constraints, then the lower mass limits of gluino and stop rise to 4 TeV and 3 TeV respectively. It is reported in \cite{Sirunyan:2020ztc} that @ 13 TeV (137 $\rm fb^{-1}$) the mass limits for the gluino and stop are about 2.1 TeV and 0.9 TeV respectively.  For the LHC Run-3, namely LHC @ 13-14 TeV (300 $\rm fb^{-1}$) the 95 $\%$ CL limit for gluino mass is about 2.4 TeV, and with 3000 $\rm fb^{-1}$ the 95 $\%$ CL limit is about 2.9 TeV \cite{Ruhr:2016xsg}. Similarly at the LHC Run-3 @ 13 TeV (300 $\rm fb^{-1}$) the 95 $\%$ CL limit for stop mass is about 2.4 TeV and rises to 2.9 TeV (3000 $\rm fb^{-1}$)  95 $\%$ CL \cite{Ruhr:2016xsg}. We also display various bounds from High-Luminosity (HL) and High-Energy @ 27 TeV (HE) LHC searches. Vertical brown, blue and yellow lines represent the reach for the stop mass at 95 $\%$ CL HL-LHC, @ 27 TeV (15 $\rm ab^{-1}$), 95$\%$ CL @ 27 TeV (15 $\rm ab^{-1}$) respectively. The horizontal purple, black and cyan lines represent for the reach for the gluino mass with 5$\sigma$ @ 14 TeV(3 $\rm ab^{-1}$), 5$\sigma$ @ 27 TeV(15 $\rm ab^{-1}$) and 95$\%$ CL @ 27 TeV(15 $\rm ab^{-1}$) respectively \cite{CidVidal:2018eel}. We see that these LHC searches may probe nearly half of the parameter space shown in green color and some part of the red points shown in the plots. 

As far as future searches at @ 100 TeV proton-proton collider are concerned, with 3000 $\rm fb^{-1}$ the gluino discovery limit is about 11 TeV \cite{Cohen:2013xda}. If the stop decays to higgsinos, it can be discovered (excluded) up to 6 (7) TeV with 3 $\rm ab^{-1}$. Moreover if the stop decays through gluinos to LSPs, one needs 30 $\rm ab^{-1}$ integrated luminosity due to additional SUSY backgrounds from gluino pair production \cite{Fan:2017rse} for a discovery.

{The top right panel of Figure \ref{gluino-stop} displays the results for the squark masses in the $m_{\tilde q}-m_{\tilde g}$ plane. Even though $m_{\tilde{q}}$ can lie in a wide range from 2 to 10 TeV (green), the constraint on the relic density of the LSP neutralino (red) shrinks this range to [4, 10] TeV. When the 300 fb$^{-1}$ integrated luminosity is collected in the collider experiments with 14 TeV center of mass energy, the squarks can be probed up to 2.7 TeV, while this scale is expected to raise to about 3 TeV when the integrated luminosity reaches to 3000 fb$^{-1}$ \cite{Cohen:2013xda}. The probing scale for the squarks will be about 6.6 TeV in the collisions of 33 TeV center of mass energy, and thus, nearly half of our solutions will likely be tested in near future collider experiments. In addition to it, it is shown in \cite{Golling:2016gvc} with integrated luminosity of $3 \,\rm{ab}^{-1}$ and $30 \,\rm{ab}^{-1}$ @ 100 TeV collider the 5$\sigma$ discovery reach is about 16 TeV and 20 TeV thus covering entire model parameter space.}


In the bottom panel, we see that the MSSM higgsino mass parameter $|\mu|$ varies from $\sim$ 0.15 TeV to 4 TeV (for green points), and {the DM constraint bounds it from below at about 2.8 TeV (red)}. We also note that the solutions with $|\mu| \sim$ 0.2 TeV are allowed since the neutralino mass in this case is heavier than 100 GeV (see Eq.~\ref{lhc-b}). As mentioned earlier, the $\mu$ term in this model,  arises from $\lambda_{\mu} SH_{u} H_{d}$ term. We expect that we may avoid little hierarchy problem ~\cite{Dimopoulos:2014aua,Cohen:2015ala,Nelson:2015cea,Martin:2015eca} with a relatively light $\mu$ term, {since Light Higgsions are required to address the electroweak fine tuning problem. Indeed, this resolution to the $\mu-$problem is also accommodated in Next to MSSM (NMSSM) \cite{R5,A5,A6,A7}. However, the existence of a $Z_{3}$ symmetry and its spontaneous breaking lead to the domain wall problem in NMSSM (See \cite{Dvali:1997uq} for possible solutions to $\mu$-problem)}.


\begin{figure}[htp!]
\centering
\subfiguretopcaptrue

\subfigure{
\includegraphics[totalheight=5.5cm,width=7.cm]{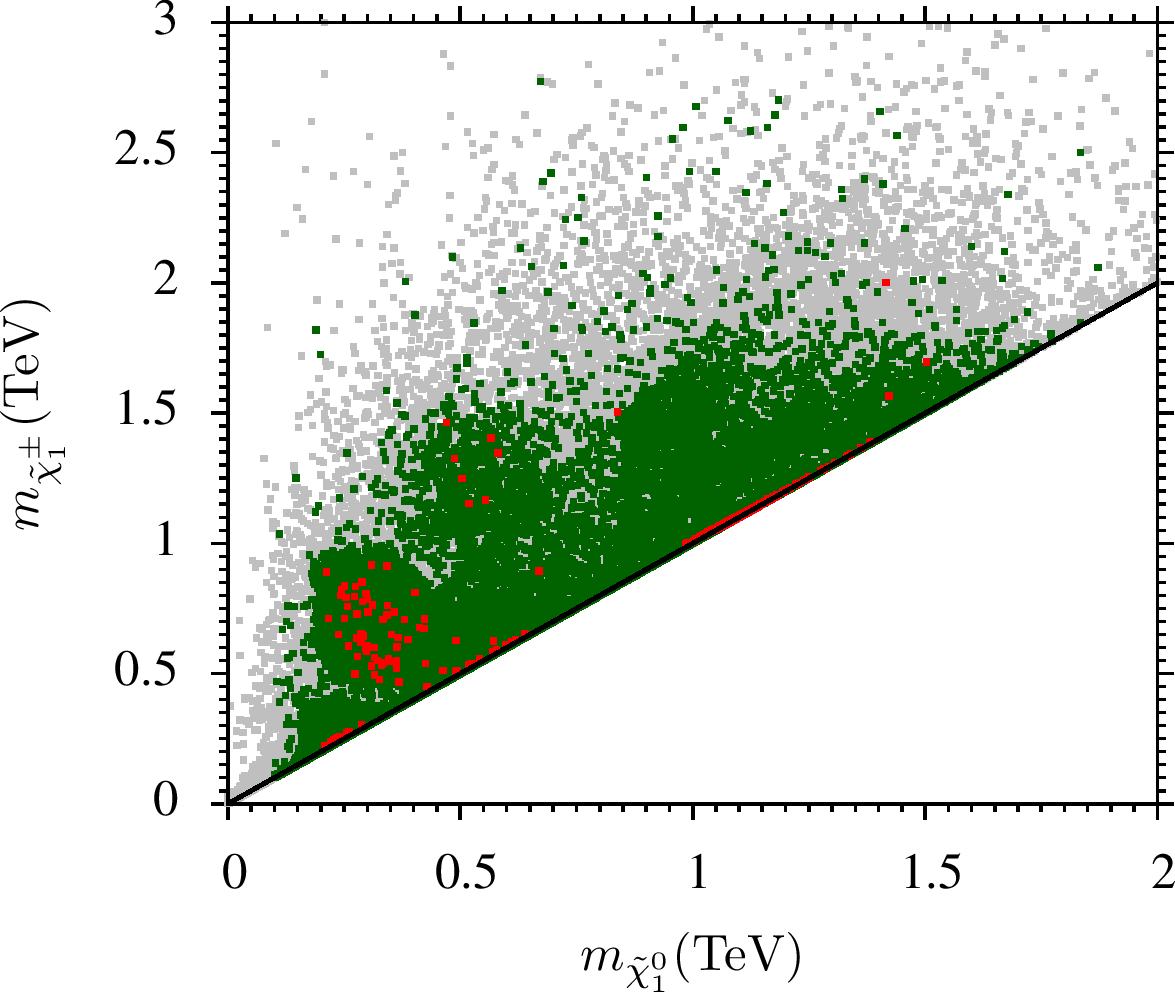}
}
\subfigure{
\includegraphics[totalheight=5.5cm,width=7.cm]{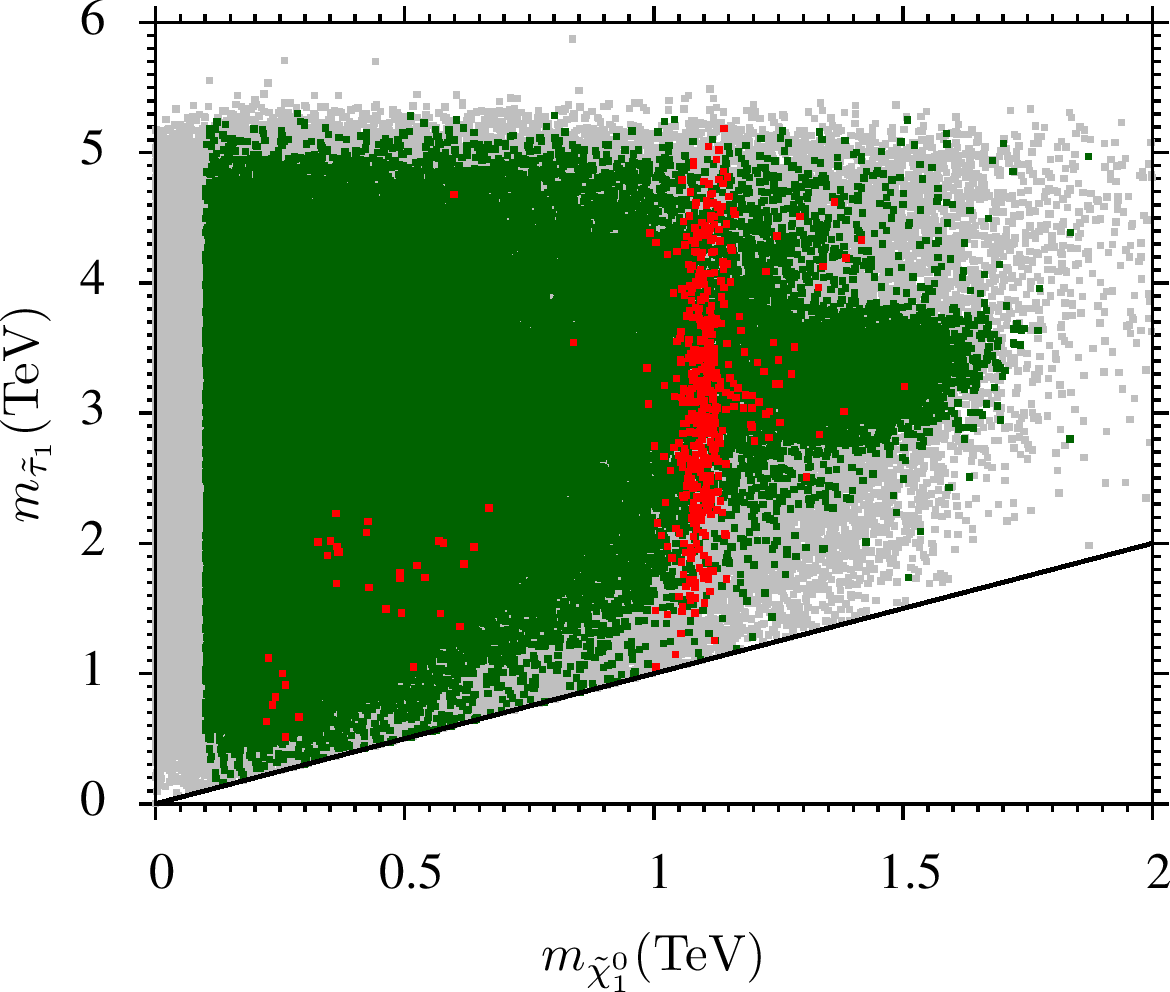}
}
\subfigure{
\includegraphics[totalheight=5.5cm,width=7.cm]{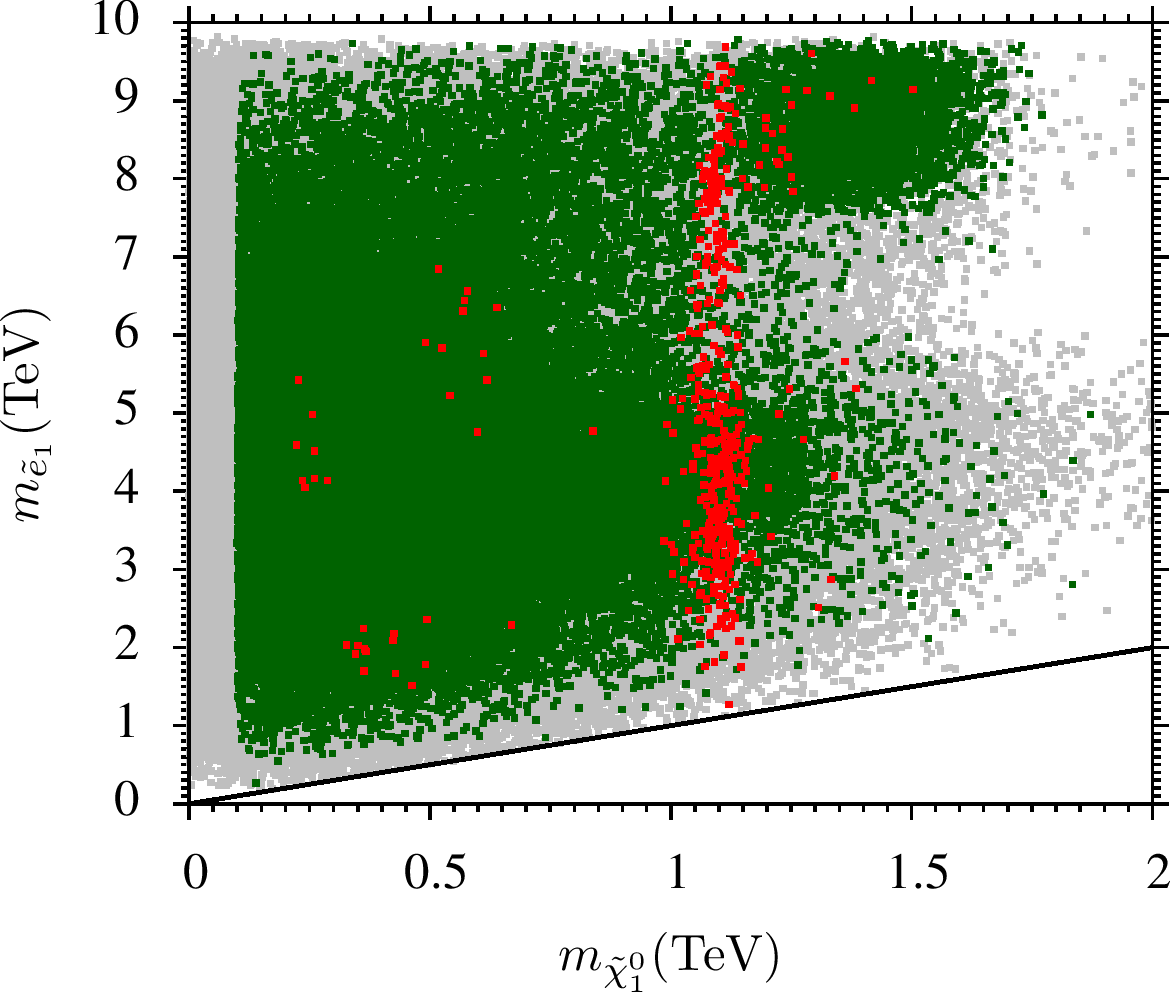}
}
\subfigure{
\includegraphics[totalheight=5.5cm,width=7.cm]{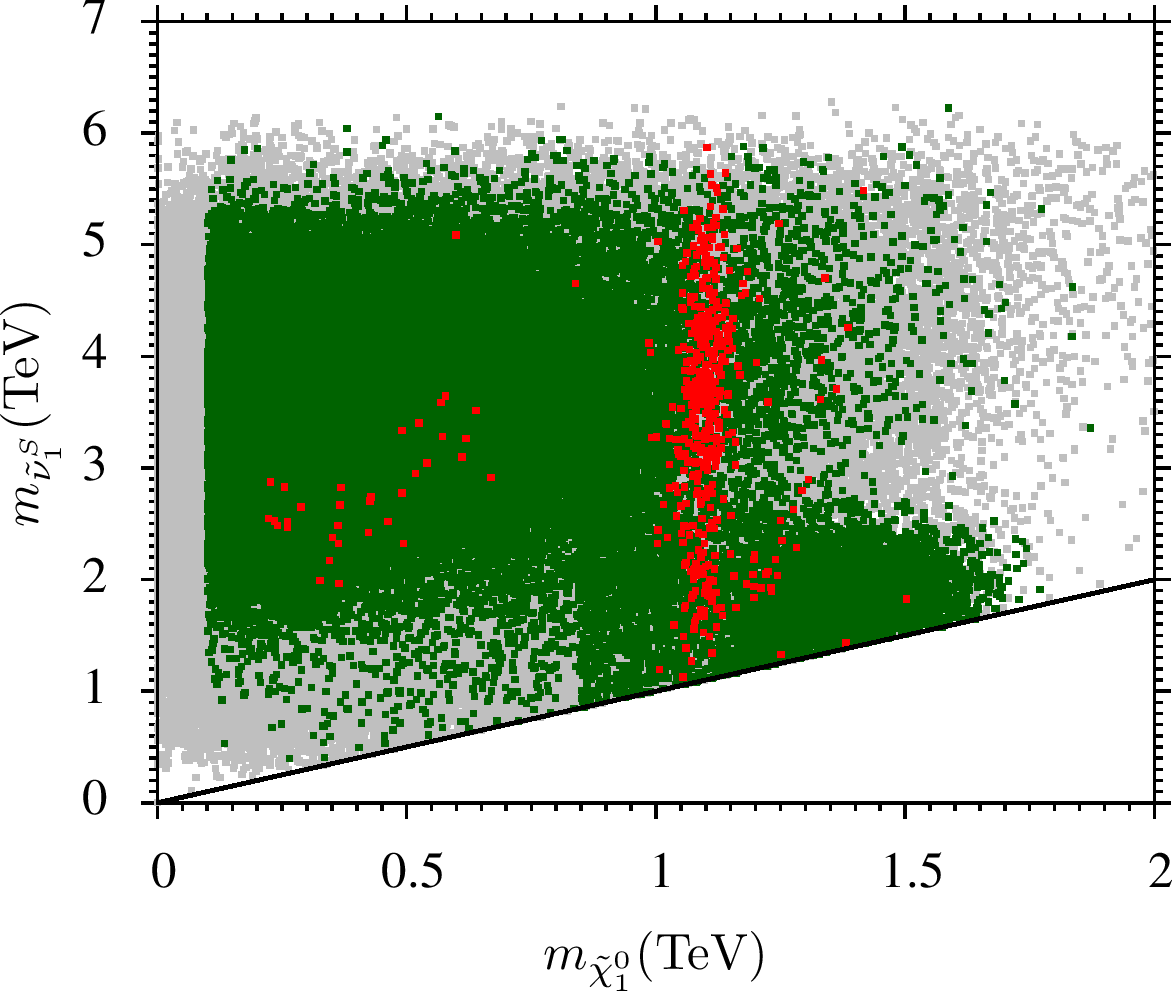}
}
\subfigure{
\includegraphics[totalheight=5.5cm,width=7.cm]{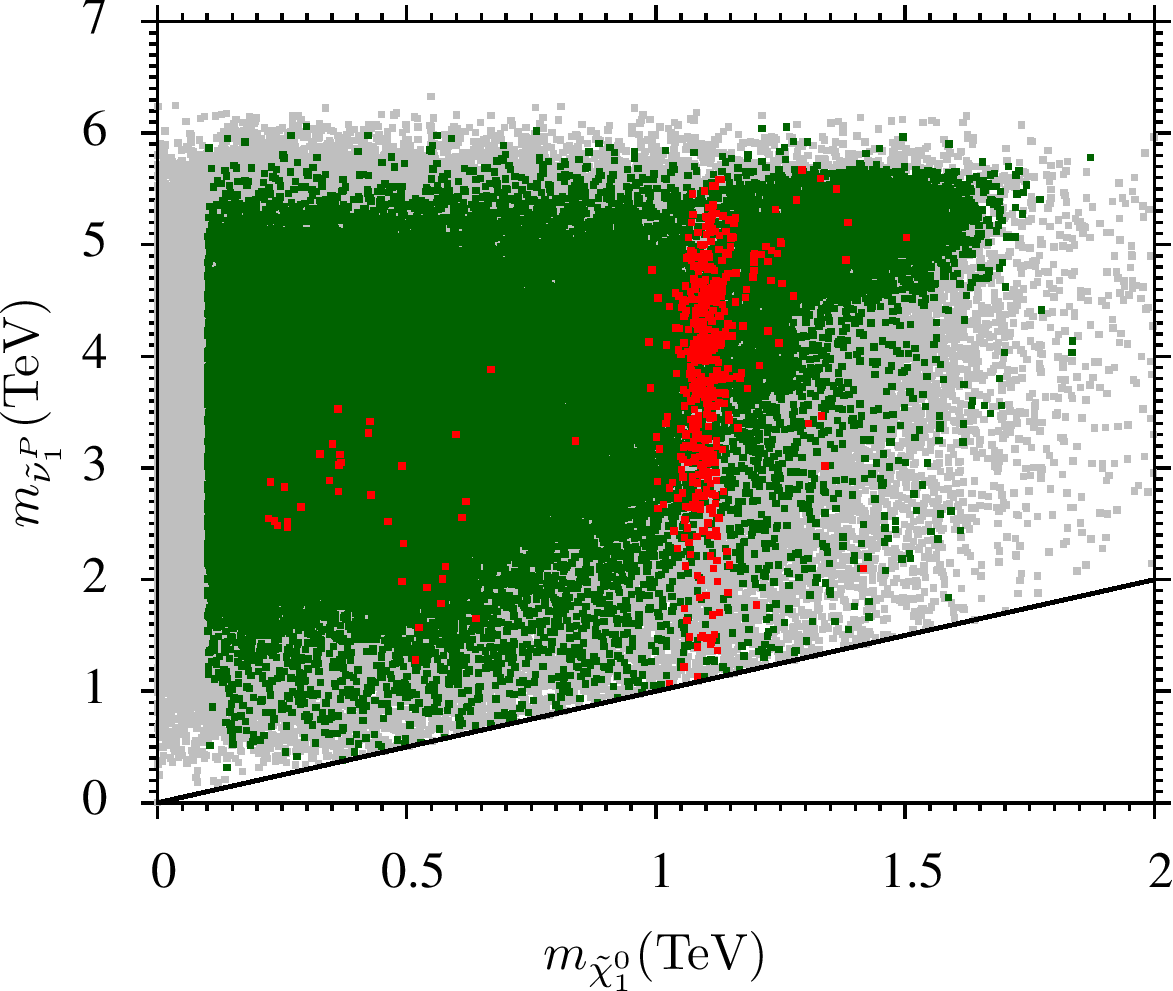}
}
\caption{Plots in the $m_{\tilde{\chi}_{1}^{\pm}}-m_{\tilde{\chi}_{1}^{0}}$, $m_{\tilde{\tau}_{1}}-m_{\tilde{\chi}_{1}^{0}}$, $m_{\tilde{e}_{1}}-m_{\tilde{\chi}_{1}^{0}}$, $m_{\tilde{\nu}^{S}_{1}}-m_{\tilde{\chi}_{1}^{0}}$ and $m_{\tilde{\nu}^{P}_{1}}-m_{\tilde{\chi}_{1}^{0}}$ planes. The color coding is the same as in Figure \ref{gluino-stop}. The diagonal lines represent the solutions with degenerate masses of the particles displayed.}

\label{sparticle}
\end{figure}

In Figure~\ref{sparticle}, we present the plots in the $m_{\tilde{\chi}_{1}^{\pm}}-m_{\tilde{\chi}_{1}^{0}}$, $m_{\tilde{\tau}_{1}}-m_{\tilde{\chi}_{1}^{0}}$, $m_{\tilde{e}_{1}}-m_{\tilde{\chi}_{1}^{0}}$, $m_{\tilde{\nu}^{S}_{1}}-m_{\tilde{\chi}_{1}^{0}}$ and $m_{\tilde{\nu}^{P}_{1}}-m_{\tilde{\chi}_{1}^{0}}$ planes. The color coding is the same as in Figure \ref{gluino-stop}. The diagonal lines represent solutions with degenerate masses of the particles displayed. These plots identify the coannihilation channels of the LSP neutralino with appropriate sparticles to yield a consistent thermal relic density. Even though it is possible to realize the lightest chargino mass at around 2.0 TeV, the results in the $m_{\tilde{\chi}_{1}^{\pm}}-m_{\tilde{\chi}_{1}^{0}}$ plane show that one can identify the chargino-neutralino coannihilation processes for $ 0.24 \lesssim m_{\tilde{\chi}_{1}^{0}}\approx m_{\tilde{\chi}_{1}^{\pm}} \lesssim 1.5$ TeV (red points around the diagonal line). A recent analyses has revealed new mass bounds on the charginos based on their decay modes. If the chargino is allowed to decay into a stau along with a suitable SM particle, then the solutions with $m_{\tilde{\chi}_{i}^{\pm}}\lesssim 1.1$ TeV are excluded \cite{Sirunyan:2017lae}, where $i=1,2$ stand for the lightest and heaviest charginos respectively. On the other hand, if the staus are heavy and the charginos cannot decay into them, then the mass bound is less restrictive, namely $m_{\tilde{\chi}_{i}^{\pm}} \geq 500$ GeV \cite{Sirunyan:2017zss}. If the mass difference between the chargino and LSP neutralino is less than the masses of the $W-$boson and charged Higgs boson, then these bounds on the chargino mass is reduced to those from LEP2 results (i.e. $m_{\tilde{\chi}_{1}^{\pm}} \gtrsim 0.1$ TeV). The approximate mass degeneracy between the LSP neutralino and the lightest chargino is one of the characteristics features of higgsino-like DM, which can be probed up to about 1 TeV \cite{Low:2014cba}. 

The $m_{\tilde{\tau}_{1}}-m_{\tilde{\chi}_{1}^{0}}$ plane represents our solutions for the stau and LSP neutralino masses. It is seen that one can realize $m_{\tilde{\tau}_{1}}\approx m_{\tilde{\chi}_{1}^{0}}$ from 0.1 TeV to about 1.3 TeV (green points); however, a significant portion of these solutions are excluded by the PLanck 2018 bound on the relic abundance of the LSP neutralino. The relic density constraint restricts the stau mass range as  $0.9\lesssim m_{\tilde{\tau}_{1}} \lesssim 1.1$ TeV, in which $m_{\tilde{\tau}_{1}}\sim m_{\tilde{\chi}_{1}^{0}}$, and the correct relic density is achieved through the stau-neutralino coannihilation processes. Interestingly, it is also possible in this region that the selectron mass is nearly degenerate with the neutralino mass (also with smuon since $m_{\tilde{e}}\approx m_{\tilde{\mu}}$). Such solutions also favor selectron(smuon)-neutralino coannihilation processes, and they can be identified in the regions with relatively small $m_{0}$ and $A_{0}$ values, which also yield $m_{\tilde{e}}\approx m_{\tilde{\mu}}\sim m_{\tilde{\tau}}$ (also see Point 3 in Table \ref{table2}). 

{A dedicated search around these points can generate more points in these regions. In an analysis \cite{Sirunyan:2018nwe}, the CMS collaboration interpreted the data in the context of simplified SUSY models such that the sleptons can be probed up to about 450 GeV in the massless LSP cases. This bound slightly decreases to 400 GeV, if the slepton is only left-handed. The significant reduction in the bound happens when the slepton is right-handed, since they do not participate $SU(2)_{L}$ interactions, and such a slepton can be probed up to only about 290 GeV. We are optimistic that the updates from the future experiments will more efficiently probe the parameter space of our model.} The last two plots of Figure \ref{sparticle} show our results for the lightest CP-even ($\nu_{1}^{S}$) and CP-odd ($\nu_{1}^{P}$) sneutrino masses versus the LSP neutralino mass. We see that along the diagonal line for $\nu_{1}^{S}$, we have red points from $m_{\nu_{1}^{S}}\sim$ 1 to 1.4 TeV. Similarly, for $\nu_{1}^{P}$, we see red points along the line from $m_{(\nu_{1}^{P})}\sim$ 1.0 to 1.1 TeV. {Such points indicate sneutrino-neutralino coannihilation processes.}

\begin{figure}
\centering
\subfiguretopcaptrue

\subfigure{
\includegraphics[totalheight=5.5cm,width=7.cm]{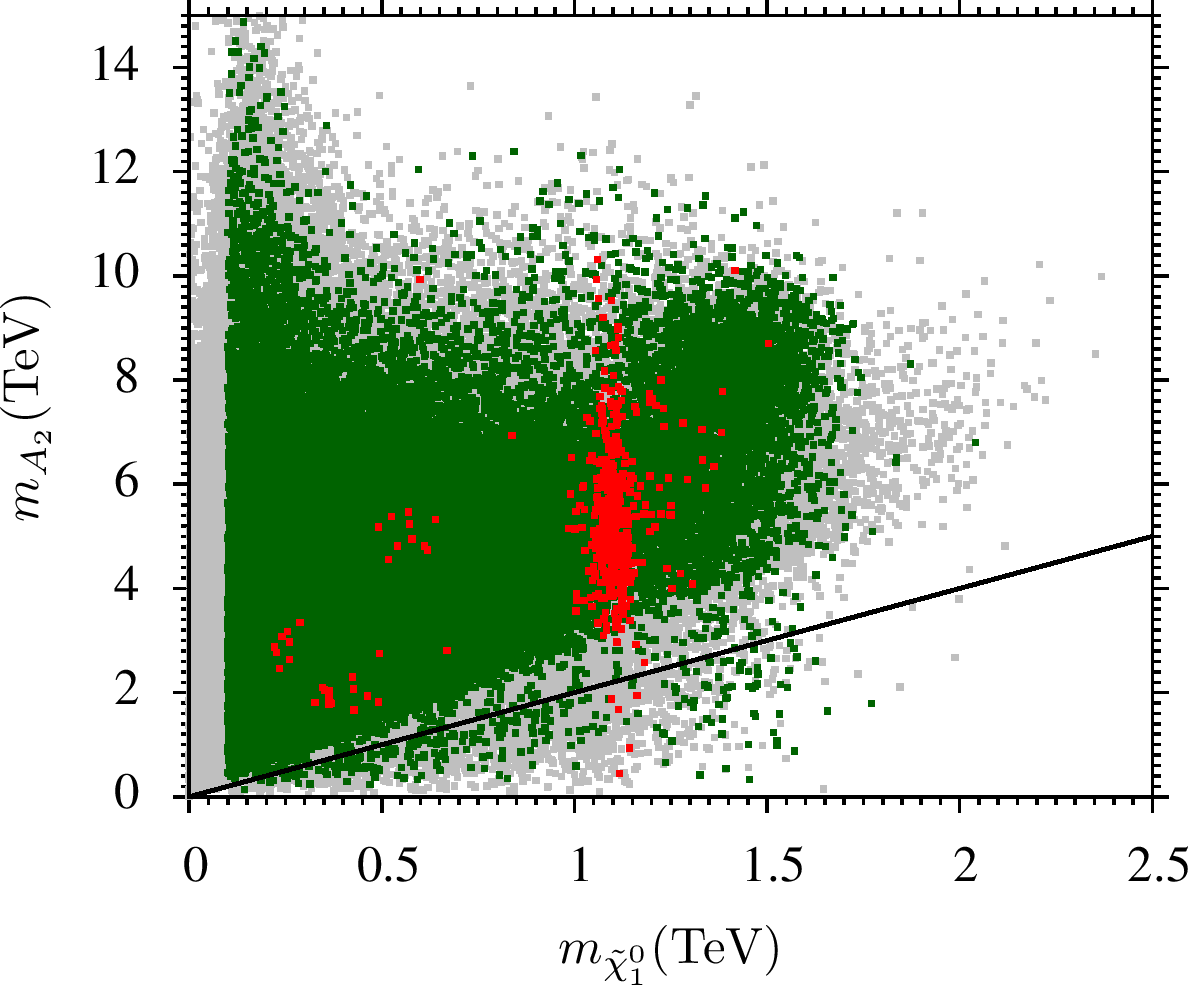}
}
\subfigure{
\includegraphics[totalheight=5.5cm,width=7.cm]{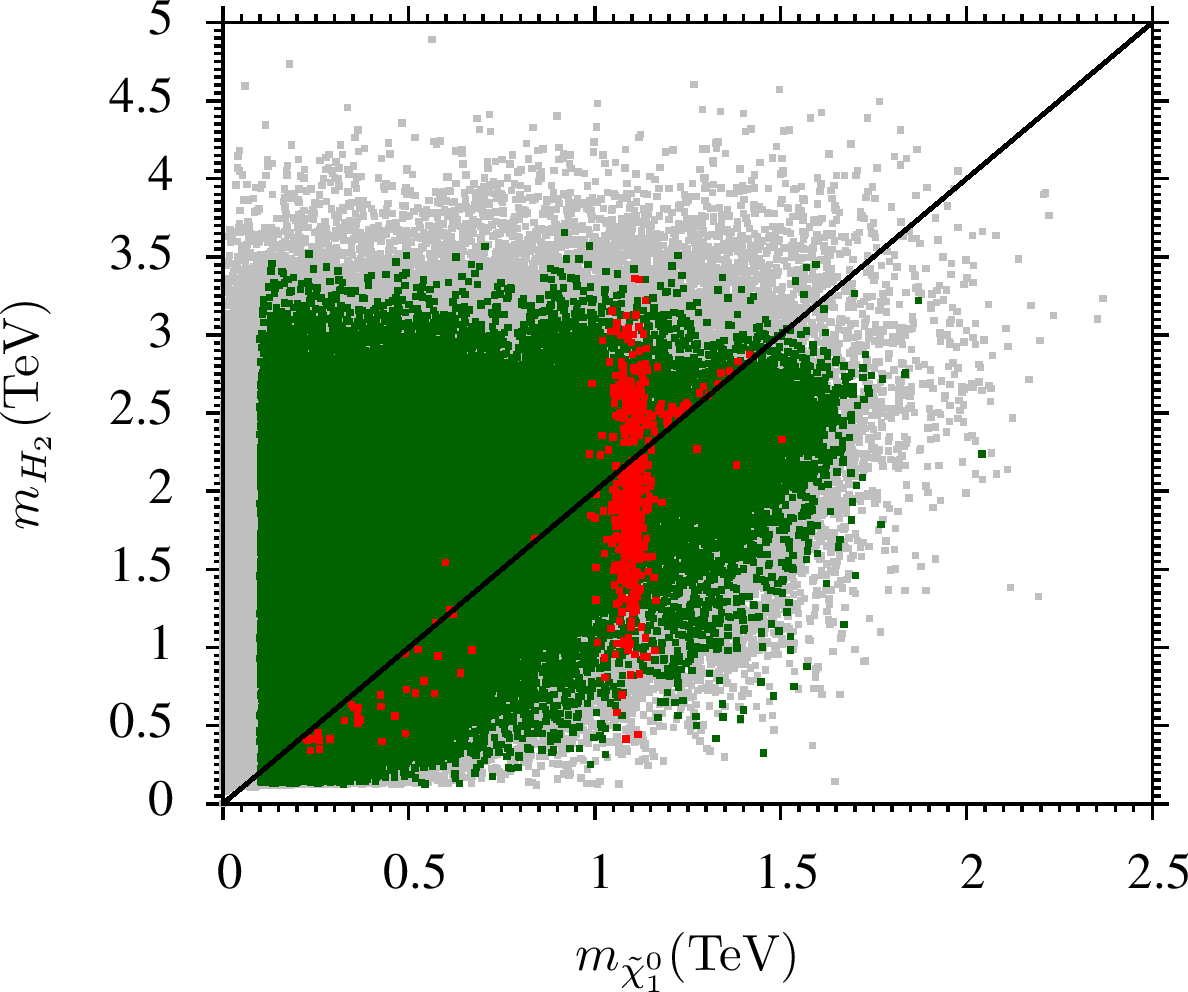}
}
\subfigure{
\includegraphics[totalheight=5.5cm,width=7.cm]{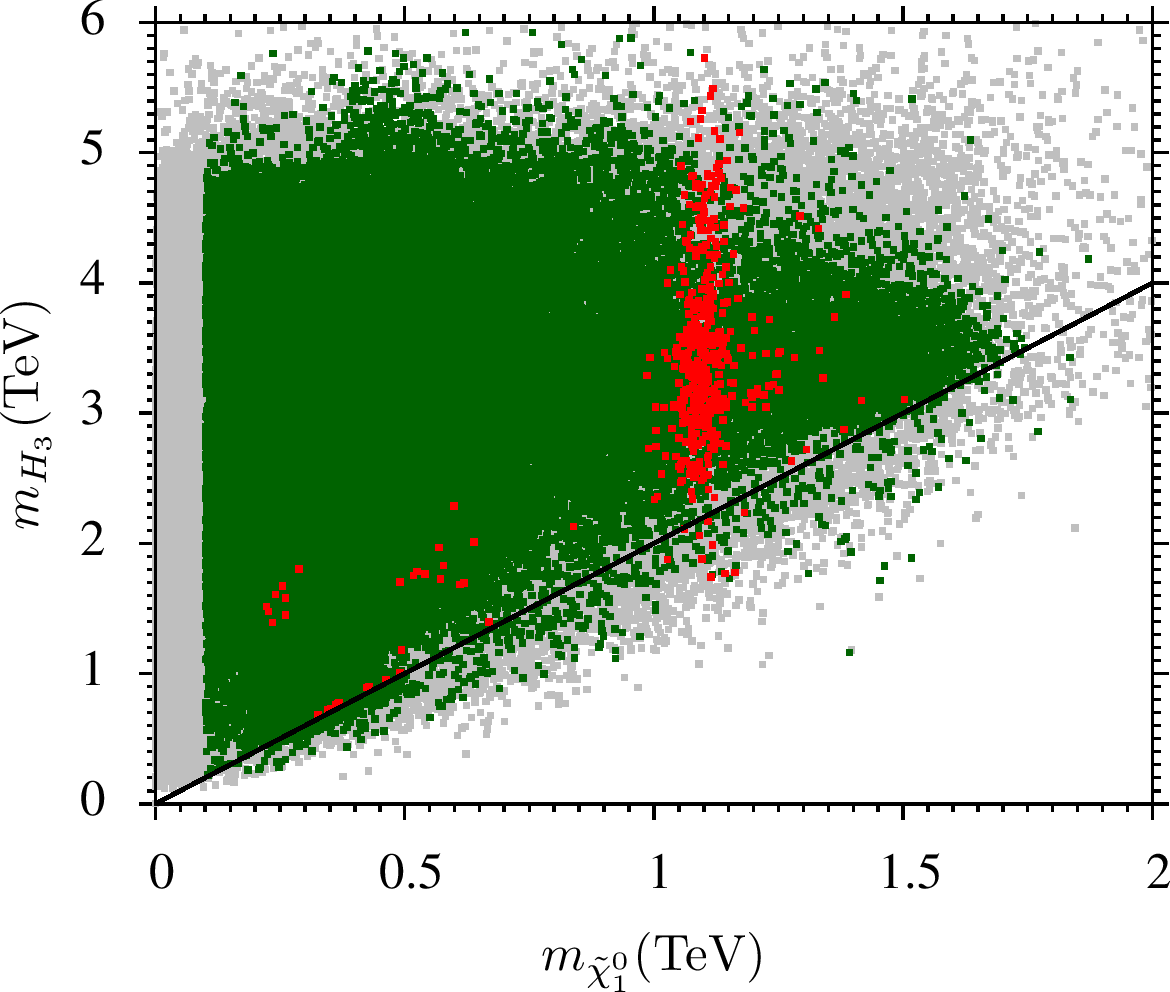}
}
\caption{Plots in the $m_{A}-m_{\tilde{\chi}_{1}^{0}}$, $m_{H_{2}}-m_{\tilde{\chi}_{1}^{0}}$ and $m_{H_{3}}-m_{\tilde{\chi}_{1}^{0}}$ planes. The color coding is the same as in Figure \ref{gluino-stop}. The diagonal lines indicate the Higgs resonance solutions for which $m_{A,H_{2},H_{3}}=2m_{\tilde{\chi}_{1}^{0}}$.}

\label{Higgs}
\end{figure}

In Figure~\ref{Higgs}, we show {the results} for various Higgs resonances with plots in the $m_{A_{2}}-m_{\tilde{\chi}_{1}^{0}}$, $m_{H_{2}}-m_{\tilde{\chi}_{1}^{0}}$ and $m_{H_{3}}-m_{\tilde{\chi}_{1}^{0}}$ planes. The color coding is the same as in Figure \ref{gluino-stop}. The diagonal lines indicate the Higgs resonance solutions for which $m_{A_{2},H_{2},H_{3}}=2m_{\tilde{\chi}_{1}^{0}}$. Even though the $m_{A_{2}}-m_{\tilde{\chi}_{1}^{0}}$ plane shows that $m_{A_{2}}$ can lie in a wide range from a few GeV to about 12 TeV (red  points), the resonance solutions with $m_{A_{2}}\approx 2m_{\tilde{\chi}_{1}^{0}}$ strictly restrict its mass as $1.1 \lesssim m_{A_{2}} \lesssim 1.3$ TeV.  The CP-even Higgs boson masses ($m_{H_{2}}$ and $m_{H_{3}}$) are bounded 3.5 TeV and 6 TeV as seen from the $m_{H_{2}}-m_{\tilde{\chi}_{1}^{0}}$ and $m_{H_{3}}-m_{\tilde{\chi}_{1}^{0}}$ planes, and the resonance solutions ($m_{H_{2,3}}\approx 2m_{\tilde{\chi}_{1}^{0}}$) constrain them further, $m_{H_{2,3}}\lesssim 2.5$ TeV. {These Higgs bosons can be constrained further over the $A/H\rightarrow \tau^{+}\tau^{-}$ processes. The current LHC analyses  result in $m_{A_{2},H_{i}} \gtrsim 1$ TeV, if these Higgs bosons decay only into a pair tau-leptons \cite{Aaboud:2017sjh,Sirunyan:2018zut}}.




%
\begin{figure}[t!]
\centering
\subfiguretopcaptrue

\subfigure{
\includegraphics[totalheight=5.5cm,width=7.cm]{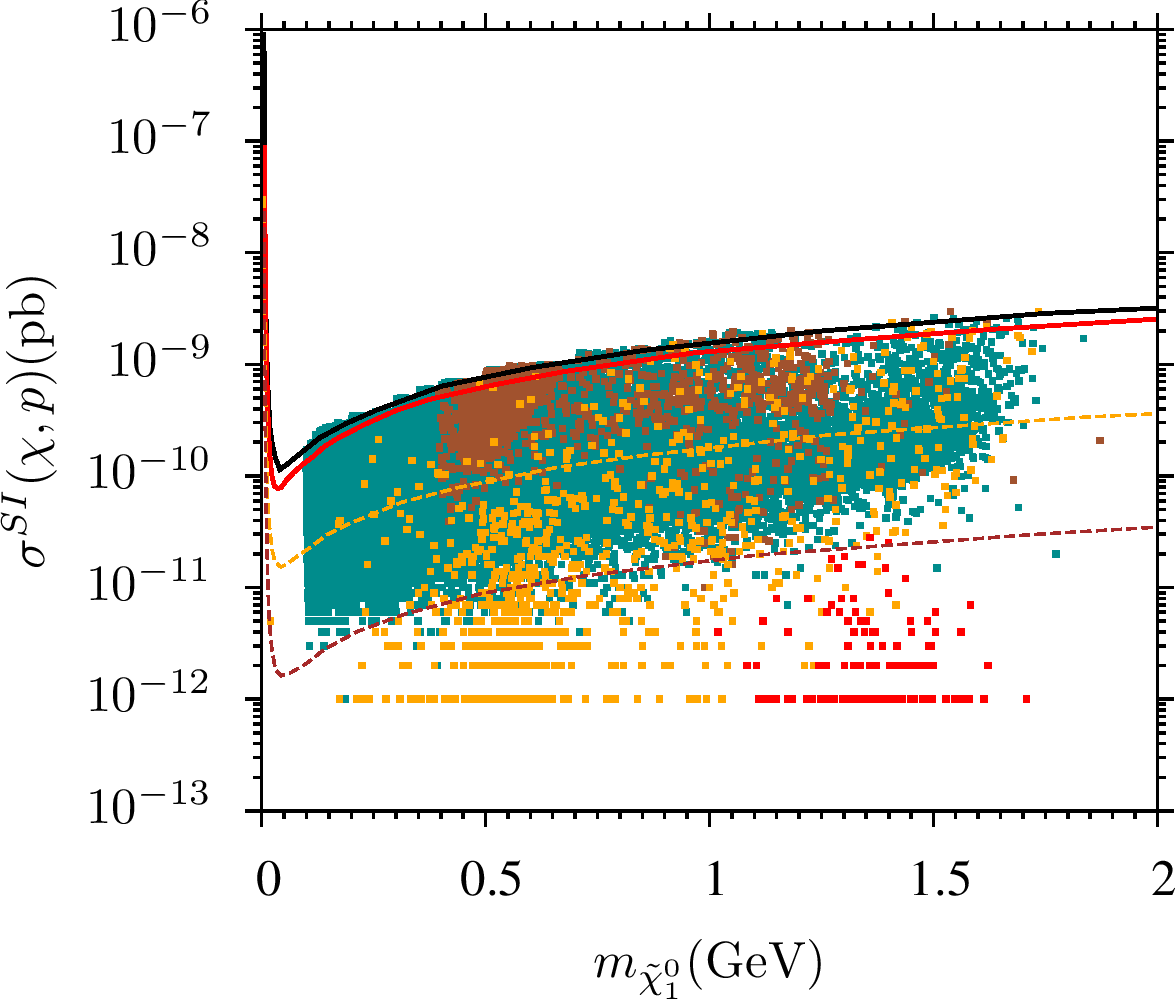}
}
\subfigure{
\includegraphics[totalheight=5.5cm,width=7.cm]{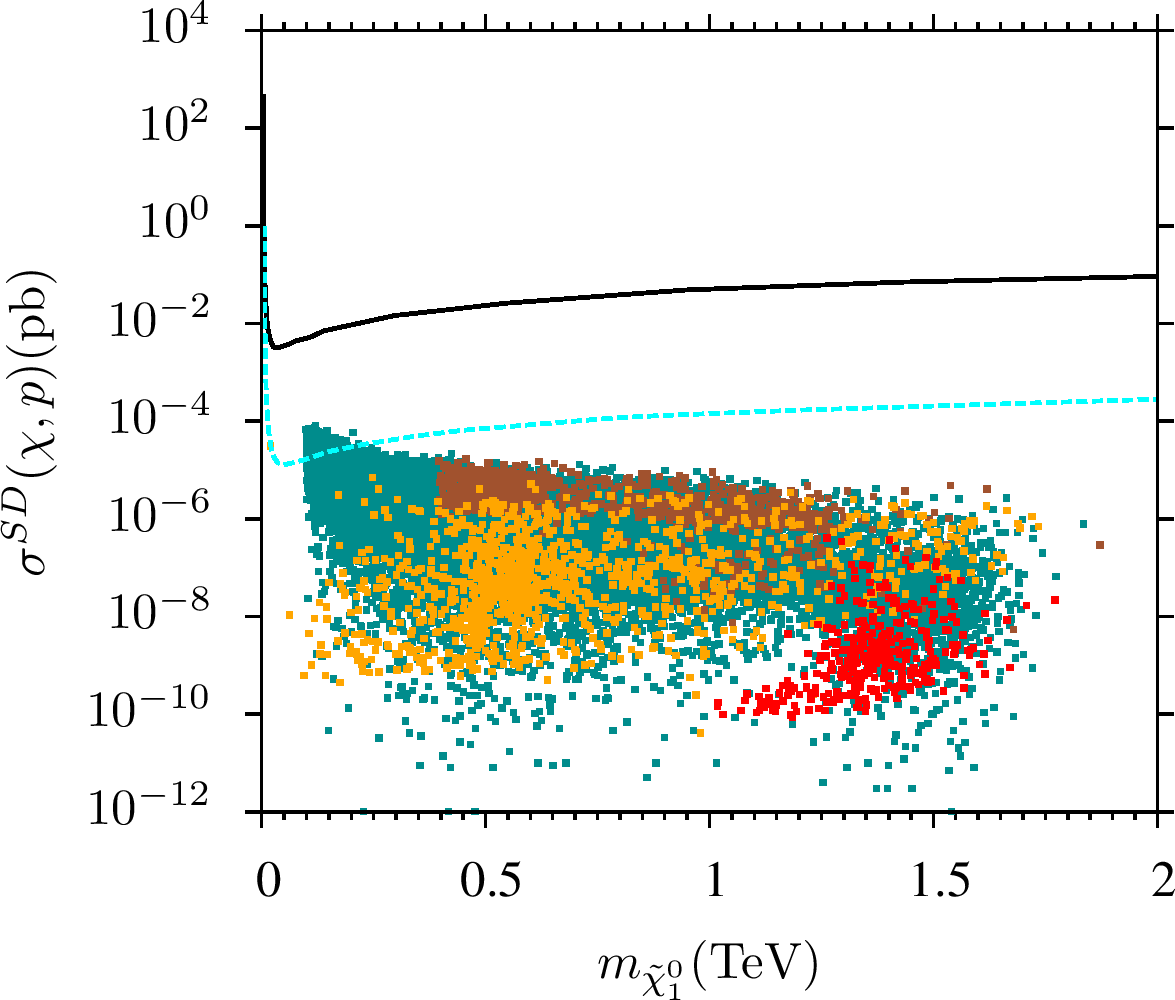}
}
\subfigure{
\includegraphics[totalheight=5.5cm,width=7.cm]{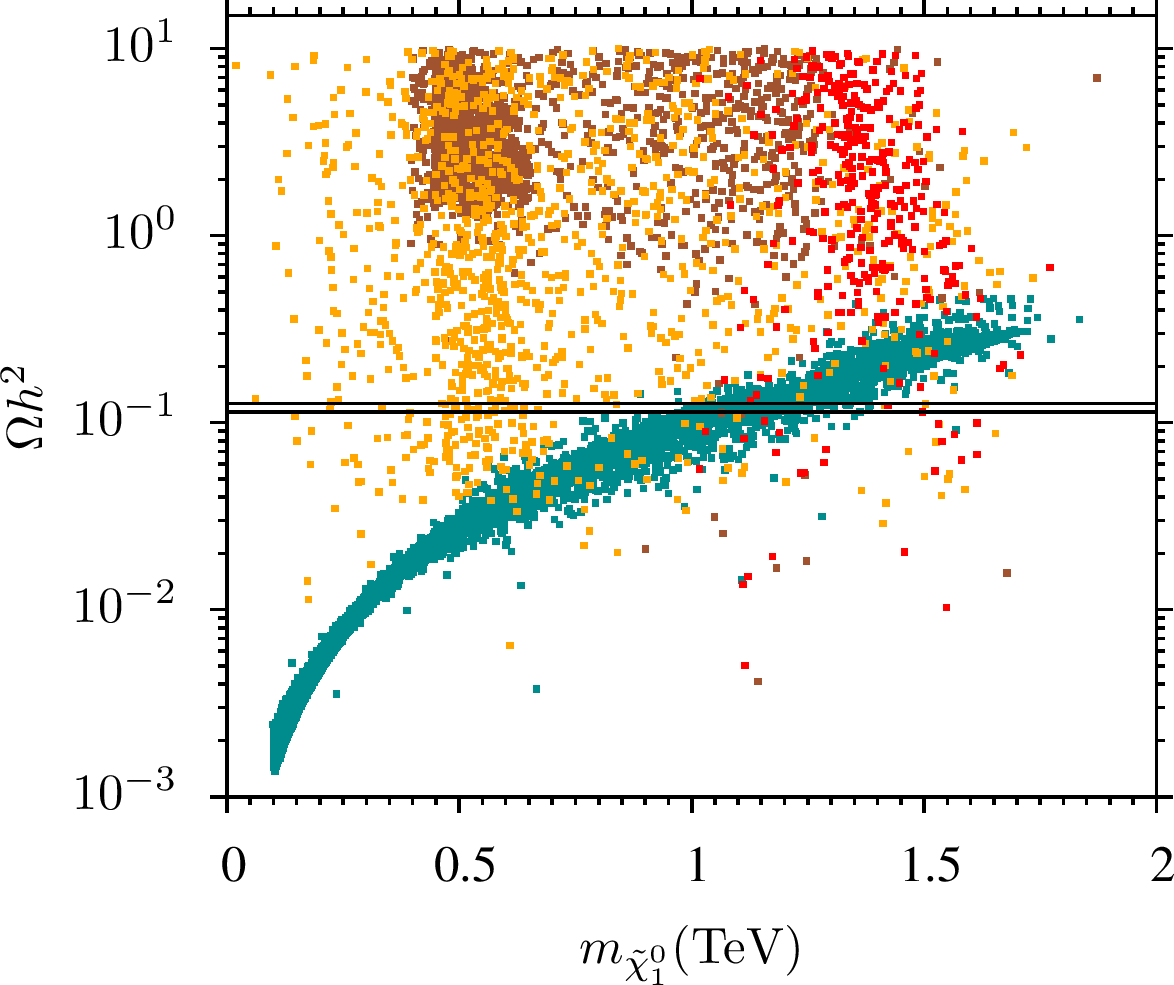}
}
\subfigure{
\includegraphics[totalheight=5.5cm,width=7.cm]{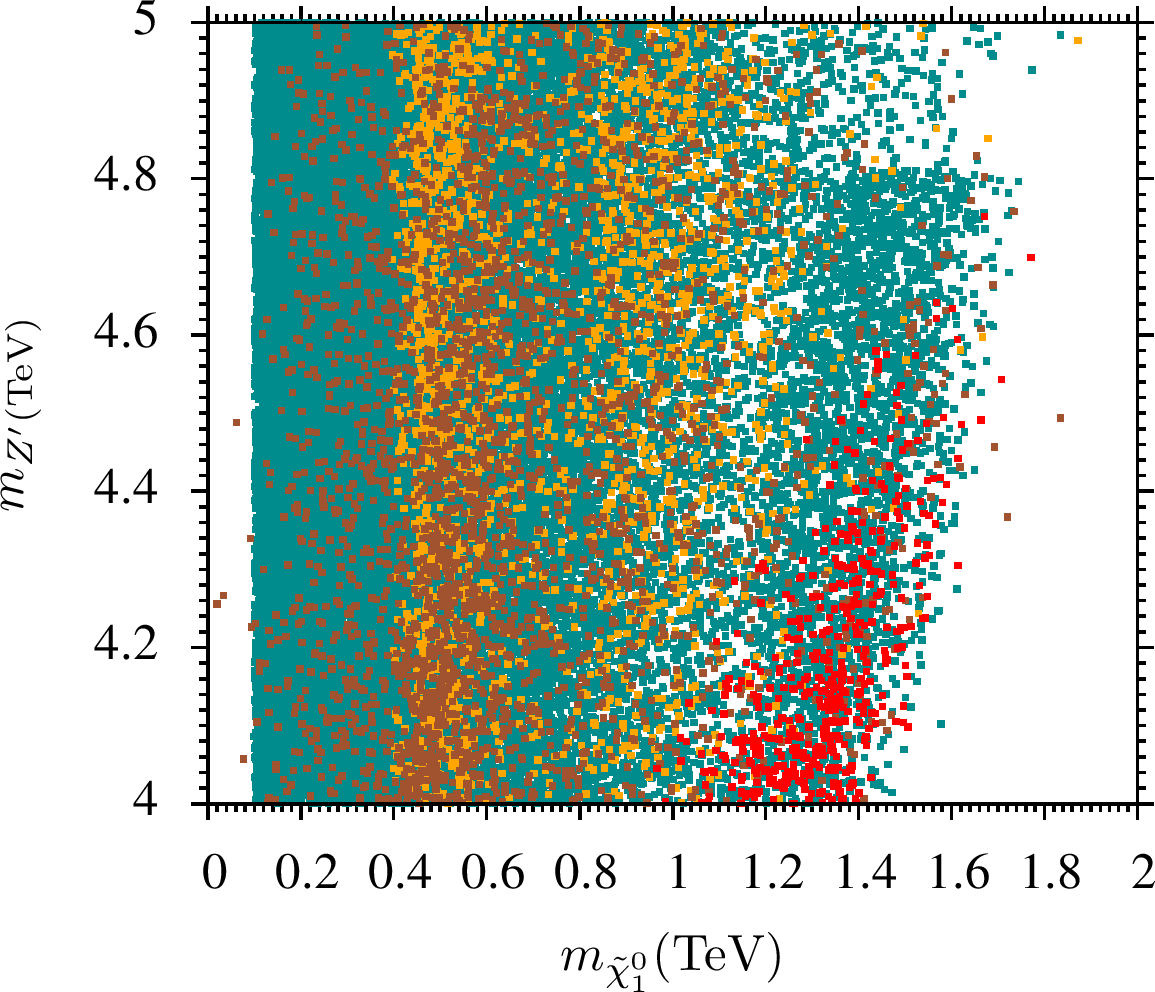}
}

\caption{Plots in the $\sigma^{{\rm SI}}-m_{\tilde{\chi}_{1}^{0}}$, $\sigma^{{\rm SD}}-m_{\tilde{\chi}_{1}^{0}}$, $\Omega h^{2}-m_{\tilde{\chi}_{1}^{0}}$ and  $M_{Z^{\prime}}-m_{\tilde{\chi}_{1}^{0}}$ planes. Turquoise points represent higgsino-type neutralino, brown points are bino-type neutralino solutions, orange points show singlino solutions, and red points represent blino-type neutralino solutions. 
The horizontal lines in the top-right panel indicate the DM relic density bounds given in Eq.(\ref{eq:omega}). In the bottom-left panel, the solid black and red lines respectively represent the current LUX \cite{Akerib:2016vxi} and XENON1T \cite{Aprile:2017iyp} bounds, and the dashed orange and brown lines show the projection of future limits \cite{Aprile:2015uzo} of XENON1T with 2 $t\cdot y$ exposure and XENONnT with 20 $t\cdot y$ exposure, respectively. In the bottom-right plot, the black solid line is the current LUX bound \cite{Akerib:2017kat} and the cyan dashed line represents the future LZ bound \cite{Akerib:2016lao}}.
\label{DM}
\end{figure}

Figure \ref{DM} display our results with plots in the $\sigma^{{\rm SI}}-m_{\tilde{\chi}_{1}^{0}}$, $\sigma^{{\rm SD}}-m_{\tilde{\chi}_{1}^{0}}$, $\Omega h^{2}-m_{\tilde{\chi}_{1}^{0}}$ and  $M_{Z^{\prime}}-m_{\tilde{\chi}_{1}^{0}}$ planes. Turquoise points represent higgsino-type neutralino, brown points are bino-type neutralino solutions, orange points show singlino solutions, and red points represent blino-type neutralino solutions. 
In the top-left panel, the solid black and red lines respectively represent the current LUX \cite{Akerib:2016vxi} and XENON1T \cite{Aprile:2017iyp} bounds, and the dashed orange and brown lines show the projection of future limits \cite{Aprile:2015uzo} of XENON1T with 2 $t\cdot y$ exposure and XENONnT with 20 $t\cdot y$ exposure, respectively. In the top-right plot, the black solid line is the current LUX bound \cite{Akerib:2017kat} and the cyan dashed line represents the future LZ bound \cite{Akerib:2016lao}. In this figure we show only those points which satisfy the direct and indirect DM bounds.
The horizontal lines in the bottom-left panel indicate the DM relic density bounds given in Eq.(\ref{eq:omega}).   The bottom-right panel shows $M_{Z'}$ versus the LSP neutralino mass.

The top panels of Figure \ref{DM} summarize our results for the direct and indirect detection experiments with respect to the LSP neuralino mass. Note that we plot only those solutions which satisfy current direct and indirect detection experimental bounds indicated above. We see that in our scans most of the solutions are higgsino-type (turquoise points) covering almost entire range of LSP neutralino mass between 100 GeV to 2 TeV. Singlino solution (orange points) are also in large number ranging neutralino mass from 0.15 TeV to 1.7 TeV. Similarly bino-type solutions are also in considerable numbers and with masses between 0.5 TeV to 1.8 TeV. Blino-type solutions are relatively small in numbers with masses between 1 TeV to 1.7 TeV. We see that current and future indirect searches will probe our solutions considerably. We also notice that blino-type solutions are small nucleon to neutralino cross section $\sigma^{{\rm SI}}$ so barely constraint even by future searches such as XENONnT with 20 $t\cdot y$ exposure. 
Similarly the top-right panel represents the spin-dependent scattering cross-sections versus the LSP neutralino mass. We see that all of our solutions are consistent with current bounds and a small portion of parameter space yielding higgsino-like DM can be probed by the future Lux-Zeplin {analyses}. However, these solutions are already excluded by the relic density constraint as discussed in the results shown in the $\Omega h^{2}-m_{\tilde{\chi}_{1}^{0}}$ plane.

{Even though a significant portion of the solutions are excluded by the DM constraint on the relic density, its impact differs for different compositions of DM.} For instance, despite the higgsino LSP solutions with $m_{\tilde{\chi}_{1}^{0}}\gtrsim 100$ GeV, the Planck2018 bound within $5\sigma$ uncertainty excludes the solutions with $m_{\tilde{\chi}_{1}^{0}} \lesssim 800$ GeV. Furthermore, the bino-like LSP solutions mostly yield a relatively large relic density, which is inconsistent with the current measurements. This is because the bino-like LSP participates in the coannihilation processes through the weak interactions, and thus its relic density cannot be adequately lowered by such processes. In contrast to the higgsino and bino-like LSP, {a singlino LSP can still be compatible with the DM constraint even if it weighs as low as about 100 GeV.} Another strong impact from the relic density constraint can be observed on the blino LSP {{whose mass}} is restricted as $m_{\tilde{\chi}_{1}^{0}}\gtrsim 1.5$ TeV. All this discussion, on the other hand, is based on the assumption that the DM density is saturated only by the LSP neutralino. If this assumption is dropped, the solutions with lower relic abundance of LSP neutralino may still be available in conjunction with other form(s) of DM \cite{Baer:2012by}. In the bottom-right plot we see that  all possible types of LSP neutralinos can be realized for almost any value of $M_{Z'}$. In this respect, the mass of $Z'$ is not very restrictive for the low scale results of our model.
\begin{table}\hspace{-1.0cm}
\centering
\begin{tabular}{|c|ccccc|}
\hline
\hline
                 & Point 1 & Point 2 & Point 3 & Point 4 & Point 5\\

\hline
$m_{0}$        & 2851.1 &   1114.2   & 1178.4    & 3864.9 & 2295.2\\
$M_{1/2}$        & 3837.2  &  2637.5   & 3645.5    & 4750.8 & 4560.7\\
$\tan\beta$       & 7.9521 &  13.699  & 5.7389   & 49.984 &8.114\\
$A_{0}$        & -155.28 &  -2208   & -975.8 & -4131.6 & 9107.4\\
$\tan\beta^{\prime}$ & 1.9475 &  1.759   & 1.9033   & 1.6519  &1.1389\\
$M_{Z^{\prime}}$  &4849.4 &  4554.5  & 4827.6    & 4146.6 & 4769.5\\
$L_0$    & -7103  & -1014           & -3001.5    & -7348.5& -4933.9\\
$\lambda_{\mu}$   & 0.35373  & 0.2910   & 0.4146    & 0.28002 & -0.43858\\
$A_{\lambda_{\mu}}$ &5109.2 & 3230.8 & 2038.2 & 22.357 & -5661\\
$v_{S}$       &4857 & 3990 & 4320.5 & 8416.9 & 3998.3\\
$\kappa$  &-0.48214 &0.2582 &-0.401 & 0.54949 & -0.5475\\
$A_{\kappa}$ &-1542.5 &3047.7 &-3035.1 & 8694.5 & -2204.9\\
${\kappa}*M^{2}$  &1663.8 & 105.85 &73.208 & 147.44 & 6347.9\\
\hline

$m_{H_{1}}$     & 126 &    122 &      123 &       125 &  125\\
$m_{H_{2}},m_{H_{3}}$  & 2053, 2821&  1038, 2130 & 1583, 3186  & 2169,2874     &2043, 4260\\
$m_{H_{4}},m_{H_{5}}$ & 5133, 7232&  4618, 4682 &  4854, 5217& 7287,9548 & 5236, 5782\\
$m_{A_{1}},m_{A_{2}}$   &40.2, 2662 & 6.239, 2662 & 6.8787,4105 & 25, 6993    & 28, 5213        \\
$m_{A_{3}}$  &7232& 4618  &4854  &9548 & 5783\\
$m_{H^{\pm}}$    & 7238 & 4620   & 3599      &9569  &5781\\
\hline
$m_{\tilde{\chi}^0_{1,2}}$
                 & \textcolor{red}{1106}, 1109 & \textcolor{red}{781}, 788 & \textcolor{red}{1121}, 1124    & \textcolor{red}{1382}, 1388    & \textcolor{red}{1083}, 1085 \\
$m_{\tilde{\chi}^0_{3,4}}$ &1938, 2058 &  1189, 1334 & 1797, 1830 & 1534, 1643 & 2225, 2294\\
$m_{\tilde{\chi}^0_{5,6}}$ &3257, 3406& 1775,2338 &2720, 3219 & 2487, 4225 &  3527, 4036\\
$m_{\tilde{\chi}^0_{7,8}}$ &4900,5410& 3706, 5428 & 4686, 5377 & 4510, 6667 & 4675, 5443\\

$m_{\tilde{\chi}^{\pm}_{1,2}}$
                 & \textcolor{red}{1108}, 3406  & \textcolor{red}{786}, 2338 &  \textcolor{red}{1124}, 3218    & \textcolor{red}{1387}, 4225  & \textcolor{red}{1085}, 4036\\
\hline
$m_{\tilde{g}}$  & 7925 &  5533 & 7479 &9671 & 9264\\
\hline
$m_{ \tilde{u}_{1,2}}$
                 & 7171, 7461 & 4796, 4971 & 6375, 6665  & 8904, 9286 &  8109, 8471\\
$m_{\tilde{t}_{1,2}}$
                 & 5728, 6792 & 3701, 4450  & 5152, 6103  & 66827, 7844 & 6701, 7864\\
\hline $m_{ \tilde{d}_{1,2}}$
                 & 7240, 7461 & 4887,4972 & 6453, 6666  & 8874, 9286   & 8013, 8471\\
$m_{\tilde{b}_{1,2}}$
                 & 6787, 7219 & 4438, 4831  & 6096, 6443  & 7865, 7891 & 7862,7991\\
\hline
$m_{\tilde{\nu}_{{\rm CP-even}}}$
                 & 4238 & 2590  & 3239  & \textcolor{red}{1435} & 3965\\
$m_{\tilde{\nu}_{CP-odd}}$
                 &  4246 & 2590 & 3239  & 4863 &\textcolor{red}{1134}\\
\hline
$m_{ \tilde{e}_{1,2}}$
                & 2916, 4247 & 2609, 4971 & \textcolor{red}{1268}, 3243  & 8904, 9286 & 2895, 3984  \\
$m_{\tilde{\tau}_{1,2}}$
                & 2895, 4240 & \textcolor{red}{783}, 891 &  1255, 3240 & 3013, 4103&2846, 3966 \\
\hline

$\sigma_{SI}({\rm pb})$
                & $1.33\times 10^{-10}$ & $3.23\times 10^{-10} $ & $ 2.4\times 10^{-10} $ & $ 3.85\times 10^{-10}$ & $2.6\times 10^{-11} $ \\

$\sigma_{SD}({\rm pb})$
                & $1.92\times 10^{-7}$ & $3.60\times 10^{-7}$ & $2.075\times 10^{-7}$ & $2.39\times 10^{-8} $ &
$1.11\times 10^{-7} $\\

$\Omega_{CDM}h^{2}$&  0.1178 & 0.101  & 0.12192 & 0.119 & 0.122\\

\hline

\hline
\end{tabular}
\caption{Masses are in units of GeV, $\mu>0$ {{and all points}}
satisfy the sparticle mass bounds, and B-physics constraints described in Section~\ref{sec:scan}. 
Points 1 and 2 represent chargino-neutralino coannihilation and stau NLSP respectively. Point 3 corresponds to selectron NLSP. Points 4 and 5 display examples of CP-even and CP-odd lightest sneutrinos respectively.
}
\label{table2}
\end{table}

\begin{table}\hspace{-1.0cm}
\centering
\begin{tabular}{|c|cccc|}
\hline
\hline
                 & Point 1 & Point 2 & Point 3 & Point 4\\

\hline
$m_{0}$        & 3500.9 &   1650.1  & 3093.3 & 4570.8    \\
$M_{1/2}$        & 2168.0  &  4718.8 & 4914.2 & 4578.2    \\
$\tan\beta$       & 14.203 &  6.8025 &15.35 & 51.398   \\
$A_{0}$        & -6722.8 &  -861.7  &-5360.7 & -4212.8  \\
$\tan\beta^{\prime}$ & 1.3478 &  1.5042 &1.134  & 1.6519   \\
$M_{Z^{\prime}}$  &4644.6  &  4674.8 &4797.7 & 4263.5    \\
$L_0$    & 1747.2  & -61.277        &   -2478.1 & -9722.2   \\
$\lambda_{\mu}$   & 0.2946  & -0.13297 &0.5898 & 0.29133   \\
$A_{\lambda_{\mu}}$ &5697.6 &-3467.7 & 8841& 27.574\\
$v_{S}$       &7388.3 & 9294.2 &3507.1 & 10686\\
$\kappa$  &0.4548 &0.17831 & 0.4139&0.59597\\
$A_{\kappa}$ &4576.6 &4100 &3464.9  &10609\\
${\kappa}*M^{2}$  &293.37 & 63182 &  2.658 & 1.2881\\
\hline

$m_{H_{1}}$     & 125 &    127&      122  & 122  \\
$m_{H_{2}},m_{H_{3}}$  & \textcolor{red}{2272}, 2700&  985,\textcolor{red}{1395}    & 1931, 2237 & 2149, \textcolor{red}{3196}    \\
$m_{H_{4}},m_{H_{5}}$ &6002,10286&  4789,5137&  4547, 8874 & 8680, 10481\\
$m_{A_{1}},m_{A_{2}}$   &4.3512,5185.6 & 63.397,2813 & 5.325,\textcolor{red}{2574}  & 32.877,8478       \\
$m_{A_{3}}$  & 10286 & 4789  & 8873   & 10481\\
$m_{H^{\pm}}$    & 10289 & 4788    & 8883  & 10503    \\
\hline
$m_{\tilde{\chi}^0_{1,2}}$
                 & \textcolor{red}{{1106}}, 1426 & \textcolor{red}{{669}},894    & \textcolor{red}{{1182}}, 1187 & \textcolor{red}{{1493}},1539  \\
$m_{\tilde{\chi}^0_{3,4}}$ &1433,1615 & 895,1704& 1790, 2515 & 1680, 1681\\
$m_{\tilde{\chi}^0_{5,6}}$ &1941,3026& 2406,3185 &2657,3338 & 2420, 4081\\
$m_{\tilde{\chi}^0_{7,8}}$ &3849,6178& 4167,6297&4359,6371 & 5300, 7201\\

$m_{\tilde{\chi}^{\pm}_{1,2}}$
                 & 1426,1940  & 894,4167 & \textcolor{red}{1186},4359 & 1679, 4081  \\
\hline
$m_{\tilde{g}}$  & 4767 & 9532 &  9951  & 9398\\
\hline
$m_{ \tilde{u}_{1,2}}$
                 & 5245,5353 & 8100,8550 & 8800, 9274 & 9010, 9347 \\
$m_{\tilde{t}_{1,2}}$
                 & 2924, 4307 & 6604, 7861  & 6589, 8209 & 6806, 7824  \\
\hline $m_{ \tilde{d}_{1,2}}$
                 & 5277, 5354 & 8125, 8551  & 8785, 9274 & 8976, 9347  \\
$m_{\tilde{b}_{1,2}}$
                 & 4296, 5164 & 7856,8107  & 8203,8638 & 7817, 7955 \\
\hline
$m_{\tilde{\nu}_{{\rm CP-even}}}$
                 & 1120 & 2915  & 1960 & 1630  \\
$m_{\tilde{\nu}_{{\rm CP-odd}}}$
                 &  3925 & 2773 & 4599 & 5268 \\
\hline
$m_{ \tilde{e}_{1,2}}$
                & 3985,5353 & 2289,3889 & 4663, 9274 & 9010, 9347   \\
$m_{\tilde{\tau}_{1,2}}$
                & 3325,3463 & 2272,3885 &  3468, 3631 & 3483, 4723 \\
\hline

$\sigma_{SI}({\rm pb})$
                & $1.43\times 10^{-10}$ & $1.1\times 10^{-11} $ & $ 3.78\times 10^{-10} $  &
                $2.0\times 10^{-12} $\\

$\sigma_{SD}({\rm pb})$
                & $2.26\times 10^{-7}$ & $3.96\times 10^{-8}$ & $3.183\times 10^{-8}$ &
                 $1.00\times 10^{-9}$\\

$\Omega_{CDM}h^{2}$&  0.1112 & 0.11763   & 0.123 & 0.1546 \\

\hline
\hline
\end{tabular}
\caption{{{Masses are in units of GeV, $\mu>0$ and all}}
points satisfy the sparticle mass bounds, and B-physics constraints described in Section~\ref{sec:scan}. 
Points 1 and 2 represent $H_2$ and $H_3$ resonance solutions respectively, and point 3 corresponds to selectron $A_2$-funnel solution . Point 4 displays an example of blino-type LSP neutralino. 
}
\label{table3}
\end{table}

Before concluding, we present some benchmark points in two Tables. {In Table \ref{table2}, point 1 represents chargino-neutralino coannihilation solutions}. Point 2 is a representative solution {{with NLSP stau}}. Point 3 displays an example where the selectron is almost degenerate in mass with the LSP neutralino. {Points 4 and 5 represent the coannihilation scenarios involving CP-even and CP-odd sneutrinos, respectively.} All these points are also examples of higgsino-type neutralino. {In Table \ref{table3}, point 1} displays $H_{2}$-resonance solution (bino-type neutralino). Point 2 is an example of $H_3$-funnel (singlino-bileptino mixed neutralino) solution, and Point 3 represents $A_{2}$-resonance solutions with higgsino-type neutralino. Point 4 depicts an example of blino-bileptino mixed LSP neutralino. 
   
\section{Conclusions}\label{conclusion}

We {{have discussed}} a class of SUSY models in which the MSSM gauge group is {{supplemented}} with a gauged $U(1)_{B-L}$ and a global $U(1)_{R}$ {{symmetry}} groups. In addition to the presence of a new neutral gauge boson ($Z'$) associated with the $U(1)_{B-L}$ symmetry, the particle content of {{this $B-L$ extended MSSM includes diphoton resonances and additional neutralino states}}. {{Indeed, the number of neutralinos is doubled in comparison with the MSSM and assuming the DM density is saturated by a LSP neutralino, the resulting model yields quite a rich phenomenology that we have explored in this paper.}} 

Exploring the collider implications of {{this}} class of models reveals {{that}} the stop and gluino {{are accessible}} in future experiments such as {{HL-LHC.}} {{In some cases lower bounds on the gluino and stop masses of 4 TeV and 3 TeV respectively are found after imposing the requirement that the LSP neutralino comprises the DM in the universe. Without \textbf{the DM} constraints the gluino and stop masses can as light as 2.2 TeV and 1 TeV respectively.}} The MSSM $\mu$ parameter can take values as low as $\sim$ 325 GeV, which may help in ameliorating the little hierarchy problem. The chargino masses vary between 0.1 - 2.6 TeV. In addition, imposing the relic density constraints yields a chargino mass as light as {{240 GeV}}, {{although}} it can also be {{as heavy as}} 1.5 TeV or so (chargino-neutralino coannihilation region). 


The mass spectrum in this region involves staus heavier than about 500 GeV. In this context, the solutions with a chargino mass lighter than 500 GeV do not allow the {{latter to}} decay into staus, and since the mass difference between the chargino and LSP neutralino is less than $W^{\pm}$ and charged Higgs {{boson masses}}, such solutions are still {{viable}}. In addition to staus, it is also possible to realize selectron, smuons, CP-even and CP-odd sneutrinos {{with}} masses of about 1 TeV and {{higher}}. {{The}} Higgs spectrum involves a CP-odd Higgs boson and two {{additional}} CP-even Higgs bosons whose masses can be below 500 GeV. Some of the $A_{2}$ resonance solutions with $\tan\beta \gtrsim 20$ may be tested further by the $A/H\rightarrow \tau^{+}\tau^{-}$ LHC searches.

The sparticle spectrum has interesting implications for the DM searches. The richness of the neutralino sector allows {{one}} to identify higgsino-like and bino-like DM, as well as singlino-like and blino-like DM. {{We have also identified}}, as indicated above, chargino-neutralino coannihilation processes. {{In addition}}, coannihilation processes involving stau, selectron, smuon and sneutrinos can be realized {{with}} their masses around 1 TeV and nearly degenerate with the LSP neutralino mass. {{We notice that in our present scans it is hard to get bino-like DM consistent with the relic density constraint, but it is possible to achieve higgsino, singlino and blino-like DM for various mass scales of the LSP neutralino}}. These solutions can be tested in the direct detection DM searches {{including}} LUX-Zeplin and Xenon-nT.


\section*{Acknowledgements}
The authors are thankful to Lorenzo Callibi for helpful discussions. Q.S thanks the DOE for partial support provided under grant number DE-SC 0013880. {This research was supported in part through the use of Information Technologies (IT) resources at the University of Delaware, specifically the high-performance computing resources.} CSU acknowledges the National Academic Network and Information Center (ULAKBIM) of The Scientific and Technological Research Center of Turkey (TUBITAK), High Performance and Grid Computing Center (Truba Resources) for calculation of results, in part, presented in this paper. 












\end{document}